\documentclass[aps,prd,twocolumn,notitlepage,superscriptaddress,nofootinbib]{revtex4-1}
\usepackage[utf8]{inputenc}
\usepackage[english]{babel}
\usepackage{amsmath}
\usepackage{amsfonts}
\usepackage{amssymb}
\usepackage{listings}
\usepackage{lipsum}
\usepackage{datetime}
\usepackage{graphicx}
\usepackage{mathtools}
\usepackage{mathrsfs}
\usepackage{dcolumn}
\usepackage{multirow}
\usepackage{color}   
\usepackage{hyperref}
\hypersetup{
    colorlinks=true, 
    pdfborder = {0 0 0.5 [3 3]},
    anchorcolor=black,
    citecolor=green,
    linktoc=all,    
    linktocpage=true,
    linkcolor=red,
    linkbordercolor={1 1 0},
}

\newcommand{\figname}[1]{{{Fig.~#1}}}

\begin{document}

\title{Gravitational wave signatures of ultralight vector bosons \\ from black hole superradiance}
\author{Nils Siemonsen}
\email[]{nsiemonsen@uwaterloo.ca}
\affiliation{Institut für Theoretische Physik, Eidgenössische Technische Hochschule (ETH) Zürich, 8093 Zürich, Switzerland}
\affiliation{Perimeter Institute for Theoretical Physics, Waterloo, Ontario N2L 2Y5, Canada}
\author{William E.\ East}
\affiliation{Perimeter Institute for Theoretical Physics, Waterloo, Ontario N2L 2Y5, Canada}


\date{\today}

\begin{abstract} 
    In the presence of an ultralight bosonic field, spinning black holes are
    unstable to superradiance. The rotational energy of the black hole is
    converted into a non-axisymmetric, oscillating boson cloud which dissipates
    through the emission of nearly monochromatic gravitational radiation.
    Thus, gravitational wave observations by ground- or space-based detectors 
    can be used to probe the existence of
    dark particles weakly coupled to the Standard Model.  In this
    work, we focus on massive vector bosons, which grow much faster through
    superradiance, and produce significantly stronger gravitational waves
    compared to the scalar case.  We use techniques from black hole
    perturbation theory to compute the relativistically-correct gravitational
    wave signal across the parameter space of different boson masses and black
    hole masses and spins.  This fills in a gap in the literature between
    flatspace approximations, which underestimate the gravitational wave
    amplitude in the non-relativistic limit, and overestimate it in the
    relativistic regime, and time-domain calculations, which have only covered
    a limited part of the parameter space. We also identify parameter ranges
    where overtone superradiantly unstable modes will grow faster than the
    lower frequency fundamental modes. Such cases will produce a distinct
    gravitational wave signal due to the beating of the simultaneously
    populated modes, which we compute.
\end{abstract}

\maketitle

\section{Introduction} \label{introduction}
A common feature in extensions to the Standard Model of particle physics, and
low-energy effective theories of quantum gravity, is the prediction of light
bosonic particles. Examples include massive bosons in open string
scenarios~\cite{Arvanitaki:2009fg,Jaeckel:2010ni,Arvanitaki:2010sy}, dark photons as a dark
matter candidate~\cite{Agrawal:2018vin,Essig:2013lka},
non-minimally coupled bosons as dark energy on cosmological scales~\cite{DeFelice:2016uil,DeFelice:2016yws}, and the axion as a solution to the CP problem in
QCD~\cite{Peccei:1977hh,Weinberg:1977ma}. Direct detection methods have so far
been unsuccessful in revealing the existence of such particles.  However, the
superradiant instability of black holes (BHs) provides a way to probe ultralight bosons
without assuming any specific coupling to other fields, except for the minimal
one to gravity.  Especially in light of the increasing reach of observatories
like LIGO/Virgo~\cite{Abbott:2016nmj,LIGOScientific:2018mvr}, there has been strong interest in using
gravitational wave (GW) observations to search for evidence of these particles.

A bosonic field incident on a spinning BH with real frequency $\omega_R$ and
azimuthal mode number $m$ will induce a flux of energy $\mathscr{E}$ and angular
momentum $\mathscr{L}$ across the BH horizon with ratio
$\mathscr{L}/\mathscr{E}=m/\omega_R$.  Combining the first and second laws of BH
thermodynamics implies that $\mathscr{E}-\Omega_H \mathscr{L}>0$
\cite{Bardeen:1973gs}, where $\Omega_H$ is the horizon frequency of the BH.  As
a consequence, if the \textit{superradiance condition} $0<\omega_R<m \Omega_H$
is satisfied, the energy and the angular momentum flux through the horizon
become negative, and the field will be superradiantly scattered, gaining energy
at the expense of the BH's rotational energy~\cite{Press:1972zz}.  Bosonic
fields with non-vanishing rest mass $m_\gamma$ (and hence mass parameter
$\mu:=m_\gamma/\hbar$) can form states that are gravitationally bound to the BH
and have a continuous flux across the horizon.  Bound states that satisfy the
superradiant condition grow exponentially in time, and can be characterized by a
complex frequency $\omega=:\omega_R+i\omega_I$ (where we take the convention
that $\omega_I>0$ for a growing mode).

Though the above discussion assumes an adiabatic description, this has
 been found to be a good approximation, even in the fully
non-linear scenario~\cite{East:2018glu,East:2017ovw}; the 
BH loses mass and angular momentum until it spins down to the point where
$\omega_R \approx m\Omega_H$ and the superradiance condition is saturated.
During the growth of the cloud,
the field can extract an excess of $\sim 9\%$ of the BH mass before the instability
is saturated \cite{East:2017ovw} and the BH-cloud configuration
reaches a quasi-equilibrium configuration. 

The existence of such scalar or vector bosonic clouds with an exponentially large
occupation number around spinning BHs has several potentially observable
consequences. The reduction in the rotational energy of the BH yields gaps in
the spin-mass plane, which could be investigated with direct spin and mass
measurements of a large BH population
\cite{Arvanitaki:2010sy,Davoudiasl:2019nlo,Cardoso:2018tly}. Additionally,
clouds of bosons can change the dynamical behavior, and emitted gravitational
radiation, of binary BHs, giving rise to transition resonances between cloud
field modes during the inspiral, or other tidal effects
\cite{Baumann:2018vus,Berti:2019wnn,Zhang:2018kib,Zhang:2019eid}. Finally, the
boson cloud will oscillate and produce gravitational radiation, even in its
saturated state\footnote{An even set of minimally coupled bosonic fields can be
arranged to form an axisymmetric energy-momentum distribution, and hence a
stationary hairy BH at the saturation point of the instability
\cite{Herdeiro:2016tmi,East:2017ovw,Herdeiro:2017phl}, though these will still
be unstable to higher $m$ superradiant modes~\cite{Ganchev:2017uuo}.}.  Hence,
the cloud slowly dissipates through emitting GWs. These sources might be
resolved individually~\cite{Arvanitaki:2014wva,Arvanitaki:2016qwi}, or
contribute to a GW stochastic background
\cite{Tsukada:2018mbp,Brito:2017wnc,Brito:2017zvb}.  The nearly monochromatic
gravitational radiation roughly falls into the LIGO band for boson masses
$m_\gamma\in (10^{-11},10^{-14})$ eV and the LISA band for boson masses
$m_\gamma\in (10^{-16},10^{-20})$ eV, and could be detected directly given
detailed predictions across the possible parameter space.

Most studies so far have concentrated on the scalar case -- i.e., a spin-0 boson --
in part due to technical difficulties associated with tackling the vector case,
which we discuss below.  However, vector bosons can actually be better
constrained by observations as their superradiant growth rates and the resulting
gravitational wave luminosities are orders of magnitude higher than scalars.
In this work, we provide detailed predictions for the gravitational signals
originating from a vector boson cloud around a spinning BH arising from
superradiance that can be used in performing searches/placing bounds using
upcoming GW observations.

Bound modes for a massive vector (or Proca) field on a Kerr spacetime
are characterized by a time-dependence with harmonic form, i.e., $e^{-i\omega
t}$ with $\omega_R<\mu$, and $\omega_I$ quantifying the change in amplitude over time
due to the dissipative boundary conditions at the horizon.
Hence, a quasi-normal field mode with azimuthal  number $m$ is superradiant 
when $0<\omega_R<m\Omega_H$ and $0<\omega_I$ are satisfied. In the
non-relativistic limit, $\mu M\ll 1$, the Proca field equations on a fixed Kerr
spacetime with mass $M$ reduce to a set of Schrödinger-type equations, which are readily
solved \cite{Baryakhtar:2017ngi}. In this limit, the real frequency is given by 
\begin{align} 
    \frac{\omega_R}{\mu}\approx 1-\frac{\mu^2
    M^2}{2(|m|+\hat{n}+S+1)^2}+\mathcal{O}((\mu M)^4)
    \label{freqguesses}
\end{align} 
and depends on the azimuthal mode $m$, overtone $\hat{n}$, and polarization
state $S\in\{-1,0,1\}$ of the massive vector field.  The imaginary part of the
frequency can be found by performing a matching procedure at the BH
horizon~\cite{Baryakhtar:2017ngi,Baumann:2018vus,Baumann:2019eav}.
Solving the linear, but fully relativistic, mode equations, however, has been a
challenge due to the fact that these equations do not decouple in the usual
Teukolsky formalism.  Alternative approaches include expanding in the limit of low
BH spin~\cite{Pani:2012bp}, utilizing time domain
calculations~\cite{Witek:2012tr,East:2017mrj}, or numerically solving the
coupled elliptic differential equations for the superradiant
modes~\cite{Cardoso:2018tly}. Recently, however, Frolov \textit{et}
al.~\cite{Frolov:2018ezx} (henceforth FKKS), have extended the ansatz found in Ref.~
\cite{Lunin:2017drx} in the massless case (see also Ref.~\cite{Krtous:2018bvk}), to 
massive vector fields. Based on the
conformal Killing-Yano 2-form, this ansatz separates the Proca quasi-normal mode
equations on the Kerr-NUT-(A)dS family of spacetimes.  We use this approach
here, in which finding the spectrum of Proca modes reduces to solving two
decoupled, differential eigenvalue problems, to systematically cover the
parameter space of superradiantly unstable modes.  One of the new results
presented here is that, for superradiantly unstable modes with $m>1$, overtone
modes ($\hat{n}>0$) can have faster growth rates (but greater $\omega_R$
values) than the fundamental ($\hat{n}=0$) modes. This occurs for near-extremal
spins in the relativistic regime for $m=2$, moderately high spins for $m=3$,
and even for non-relativistic values for $m\geq 4$.  In such cases, the
overtone modes will reach saturation first, and then begin to decay as the
fundamental mode grows, leading to a unique GW signal with multiple frequency
components, including a lower frequency beating, or ``transition"
signal~\cite{Baryakhtar:2017ngi}, as we detail here.

In contrast to a compact object binary, the GWs from a bosonic cloud around a
spinning BH cannot be approximated by a series of multipoles in a
post-Newtonian framework. The characteristic length scale of the bosonic
condensate $r_c$ can be comparable to, or much larger, than the wavelength of the
emitted gravitational radiation (depending on the field's mass):
$\lambda_\text{GW}\gg r_c$.  Hence, the quadrapole GW formula breaks down in
much of the parameter space.  In Ref.~\cite{Baryakhtar:2017ngi}, the GWs from
vector boson clouds were calculated in the non-relativistic limit on a flat
spacetime. However, as explained there, even in calculating the leading order
non-relativistic result, the effect of the BH spacetime must be taken into
account, and can increase the GW power by an estimated factor of $10$.
References \cite{East:2017mrj,East:2018glu} computed the GW luminosity using fully
relativistic time-domain simulations, but due to the computational expense,
they only covered a limited number of cases in the relativistic regime.  In this
work, we compute GWs using the Teukolsky formalism
\cite{Teukolsky:1973ha,Press:1973zz}, which captures linear metric
perturbations on a Kerr background, across the entire frequency spectrum. The
differential equations of this formalism are of the Sturm-Liouville type, and
can be solved using a Green's function approach.  The gravitational radiation
at future null infinity is determined by convolving this Green's function
with the source functions provided by the solutions to the Proca field
equations. Gravitational wave modes and energy flux can then be projected out
of the linear gravitational perturbations at future null infinity.  This allows
us to obtain accurate predictions for the GW amplitude and frequency across the
parameter space of Proca masses and BH masses and spins, filling in the gap in
the literature.

We also describe the evolution of vector boson clouds, assuming the superradiant
instability is triggered by a small seed perturbation---e.g.  a quantum
fluctuation, and making use of a quasi-adiabatic description (which is a good
approximation as indicated by \cite{East:2018glu,Ficarra:2018rfu}).  Generically,
the cloud will grow to saturation and then, on a longer timescale, dissipate
through GW emission.  We estimate how the frequency and amplitude of the
emitted GWs change due to the secular loss of mass and angular momentum due to
radiation. 

This paper is structured in the following way. In Sec.~\ref{kerrspacetime}, we
briefly introduce the Kerr geometry, and we review how the FKKS ansatz separates
the Proca equations on Kerr in Sec.~\ref{fieldequations}. We describe how we
solve the radial and angular equations to obtain the superradiant modes in
Sec.~\ref{angularequation}, and we detail how we normalize these modes in
Sec.~\ref{evolution}.  In Sec.~\ref{procaresults}, we present the exponentially
growing solutions to the Proca field equations. In
Secs.~\ref{teukolskyformalism} and \ref{greensfunctionsec}, we review the
Teukolsky formalism and associated Green's function method we use to calculate the
GWs.  We present results for the gravitational radiation in
Sec.~\ref{GWresults}, including the frequency and observability of GWs from ultralight vector
boson clouds in Sec.~\ref{subsec:obs}. In
Sec.~\ref{sec:discuss}, we discuss our results and conclude.  We use units with
$G=c=1$ and the $(-,+,+,+)$ metric signature throughout the paper.

\section{The Proca field solution}

\subsection{Kerr spacetime} \label{kerrspacetime}
We ignore any non-linear interactions and solve the test-field Proca equations
on a spinning BH background $g_{\mu\nu}\rightarrow g_{\mu\nu}^\text{Kerr}$
characterized by the BH  mass $M$ and spin parameter $a=J/M$ (where $J$ is 
the BH's angular momentum).
Following Ref.~\cite{Frolov:2018ezx}, the
Kerr geometry in canonical coordinates $(\tau,\zeta,y,\psi)$ takes the form
\begin{align} 
    \begin{aligned} ds^2= \frac{\Xi}{\Pi}d\zeta^2 & \
+\frac{\Pi}{\Xi}(d\tau+y^2d\psi)^2\\ & \ +\frac{\Xi
dy^2}{y^2-a^2}-\frac{y^2}{\Pi}(d\tau^2+\zeta^2d\psi)^2, 
    \end{aligned}
\end{align} 
with $\Xi:=y^2-\zeta^2$ and $\Pi:=-\zeta^2+2iM\zeta+a^2$. The canonical
coordinates are related to the usual Boyer-Lindquist coordinates by $\tau=t-a\phi$,
$\zeta=ir$, $y=a\cos\theta$, and $\psi=\phi/a$. As a member of the Kerr-NUT-(A)dS family of
spacetimes, a spinning BH spacetime has a timelike
$\boldsymbol{\xi}:=\partial_\tau$ and axial
$\boldsymbol{\eta}:=\partial_\psi$ Killing field. In addition, this
set of solutions admits more ``hidden" symmetries, which were used in
Ref.~\cite{Frolov:2018ezx} to construct an ansatz to separate the massive vector
field equation. Among others, one can construct a non-degenerate closed
conformal Killing-Yano 2-form $\boldsymbol{h}$. The tensor components of this
symplectic form are constructed from $\nabla_\alpha
h_{\beta\gamma}=2g_{\alpha[\beta}\xi_{\gamma]}$, where $\xi_\mu$ are the
components of the timelike Killing field $\boldsymbol{\xi}$. Therefore,
$\boldsymbol{h}$ takes the form 
\begin{align} 
    \boldsymbol{h}=\zeta d\zeta\wedge
(d\tau+y^2 d\psi)+y dy\wedge (d\tau+ \zeta^2 d\psi) 
\label{ckyform} 
\end{align}
in canonical coordinates, which we utilize in the next section to present the
separated Proca equation.

\subsection{Field equations} \label{fieldequations}
Disregarding any self-interactions of the massive
vector, the field equations are
\begin{align}
\nabla_\mu F^{\mu\nu}=\mu^2 A^\nu,
\label{procafieldeq}
\end{align}
where 
$F_{\mu\nu}:=2\nabla_{[\mu}A_{\nu]}$.
The non-vanishing mass $\mu$ causes an explicit breaking of the U(1) gauge
symmetry, and ensures the absence of any gauge freedom in $A_\mu$: the Lorentz
condition $\nabla_\mu A^\mu=0$ is identically satisfied.  The Kerr geometry is
Ricci-flat, and hence the equations of motion reduce to a massive vector wave
equation with mass $\mu$ in Kerr spacetime, $\nabla_\nu \nabla^\nu A_\mu=\mu^2
A_\mu$.  
The energy-momentum tensor associated with this minimally coupled massive
vector field is 
\begin{align}
\begin{aligned}
T_{\mu\nu}(A)=\mu^2 A_\mu A_\nu & +F_{\mu\alpha}F_\nu {}^\alpha \\
& -\frac{1}{4}g_{\mu\nu}\left[ F_{\alpha\beta}F^{\alpha\beta}+2\mu^2 A_\alpha A^\alpha\right].
\end{aligned}
\label{energymomentumtensor}
\end{align}
The separation ansatz for Eq.~\eqref{procafieldeq} in Boyer-Lindquist coordinates \cite{Lunin:2017drx,Krtous:2018bvk,Frolov:2018ezx} is
\begin{align}
A^\mu =B^{\mu\nu}\nabla_\nu Z, & & Z=R(r)S(\theta)e^{-i\omega t+i m \phi},
\label{fkksansatz}
\end{align}
and is formulated with the set of eigenfunctions $\{e^{-i\omega t},
e^{im\phi}\}$ and eigenvalues $\{\omega, m\}$ of the Lie derivative along the
generators of the axial and
timelike symmetries of Kerr, respectively, i.e., $L_{\boldsymbol{\xi}}$ and
$L_{\boldsymbol{\eta}}$.
The polarization tensor $B^{\mu\nu}$ of the vector field is
implicitly defined by $B^{\mu\nu}(g_{\nu\gamma}+i \nu
h_{\nu\gamma})=\delta^\mu_\gamma$, where $\nu$ is a separation constant to
be solved for.

The field equations~\eqref{procafieldeq} separate when using the FKKS
ansatz \eqref{fkksansatz}. In the Boyer-Lindquist coordinates of the background,
the radial Proca field equation takes the form
\begin{align}
\frac{d}{dr}\left[ \frac{\Delta}{q_r}\frac{dR}{dr}\right]+\left[\frac{K_r^2}{q_r\Delta}+\frac{2-q_r}{q_r^2}\frac{\sigma}{\nu}-\frac{\mu^2}{\nu^2}\right]R=0,
\label{fkksradeq}
\end{align}
whereas the angular equation is given by
\begin{align}
\frac{1}{\sin\theta}\frac{d}{d\theta}\left[ \frac{\sin \theta}{q_\theta}\frac{dS}{d\theta}\right]-\left[\frac{K_\theta^2}{q_\theta\sin^2\theta}+\frac{2-q_\theta}{q_\theta^2}\frac{\sigma}{\nu}-\frac{\mu^2}{\nu^2}\right]S=0.
\label{fkksangleq}
\end{align}
Here, we are using the following definitions
\begin{align}
\begin{aligned}
K_r:=& \ am-(a^2+r^2)\omega, & & K_\theta := m-a\omega\sin^2\theta, \\
q_r:=& \ 1+\nu^2r^2, & & \  q_\theta := 1-\nu^2a^2\cos^2\theta, \\
\sigma:= & \ a\nu^2(m-a\omega)+\omega. & &
\end{aligned}
\end{align}
The radial and angular equations are complemented by 
$L_{\boldsymbol{\xi}}^2 e^{-i\omega t}=-\omega^2 e^{-i\omega t}$ and
$L^2_{\boldsymbol{\eta}} e^{im\phi}=-m^2e^{im\phi}$. This is a system of
ordinary differential equations only coupled by the generalized eigenvalues
$\{\nu,m,\omega\}$. The boundary condition $\phi\sim\phi+2\pi$ fixes $m$ to be
an integer, while $\omega,\nu\in\mathbb{C}$. A general solution $A_\mu$, and therefore
$Z$, is constructed by summing over all eigenfunctions corresponding to
specific boundary conditions (assuming such functions form a complete set
spanning the function space of solutions). However, since we solely focus on
the superradiating quasi-normal modes, we consider only individual Proca field
modes $A_\mu^{m,\omega,\nu}$ labeled by their eigenvalues, given by the corresponding 
\begin{align}
Z\rightarrow Z^\nu_{m\omega}(t,r,\theta,\phi):=C_{m\omega}^\nu R^\nu_{m\omega}(r)S^\nu_{m\omega}(\theta)e^{-i\omega t}e^{im\phi}.
\label{zansatz}
\end{align} 
Here $C^\nu_{m\omega}$ is a normalization constant, which we will determine by specifying
the mass contained in the saturated Proca field mode around the BH (see Sec.~\ref{evolution}).

\subsection{Numerically solving the field equations} \label{angularequation}
In order to solve Eqs.~\eqref{fkksradeq} and
\eqref{fkksangleq}, we follow the approach of Ref.~\cite{Dolan:2018dqv}. We
iteratively determine the eigenvalues $\nu$ and $\omega$ of superradiating
field modes characterized by the eigenfunctions $R^\nu_{m\omega}(r)$ and
$S^\nu_{m\omega}(\theta)$ that satisfy the boundary conditions for bound
states.

Assuming regularity at the poles $\theta=0$ and $\pi$, we can expand the
solution $S^\nu_{m\omega}(\theta)$ of Eq.~\eqref{fkksangleq} in terms of 
$K_{\ell m}(\theta)$, which are the associated
Legendre polynomials modulo normalization factor: $K_{\ell m}(\theta)\propto
P^m_\ell(\cos\theta)$. Then the ansatz to solve Eq.~\eqref{fkksangleq} is given by
\begin{align}
S^\nu_{m\omega}(\theta)=\sum_{\ell'=|m|}^\infty b_{\ell'} K_{\ell' m}(\theta).
\end{align}
Plugging this ansatz into the angular equation, and
projecting out the $\ell$-modes with $\int_{-1}^1 d\cos\theta K^*_{\ell
m}(\theta)$ (here, and in the following, $*$ denotes complex conjugation), yields the non-linear eigenvalue problem:
\begin{align}
\sum_{\ell'=|m|}^\infty M_{\ell \ell'}b_{\ell'}=0 \  ,
\label{anglequation}
\end{align}
where
\begin{align}
\begin{aligned}
M_{\ell\ell'}=& \ \delta_{\ell\ell'}(\Lambda-\ell'(\ell'+1)) \\
 & \ +c^{(2)}_{\ell\ell'} a^2(\nu^2\ell'(\ell'+1)-2\sigma\nu-\nu^2\Lambda+\gamma^2) \\
 & \ -c_{\ell\ell'}^{(4)}\gamma^2a^4\nu^2-2d^{(1)}_{\ell \ell'}a^2\nu^2,
\end{aligned}
\end{align}
together with
\begin{align}
\begin{aligned}
c^{(2)}_{\ell\ell'}= & \ \frac{2\sqrt{\pi}}{3}\langle\ell,0,\ell'\rangle_m+\frac{4}{3}\sqrt{\frac{\pi}{5}}\langle\ell,2,\ell'\rangle_m,\\
c^{(4)}_{\ell\ell'}= & \ \frac{2\sqrt{\pi}}{5}\langle\ell,0,\ell'\rangle_m+\frac{8}{7}\sqrt{\frac{\pi}{5}}\langle\ell,2,\ell'\rangle_m + \frac{16\sqrt{\pi}}{105}\langle\ell,4,\ell'\rangle_m.\\
d^{(1)}_{\ell\ell'}= & \ \sqrt{\frac{4\pi}{3}}\Bigg\{\ell'\sqrt{\frac{(\ell'+1)^2-m^2}{(2\ell'+1)(2\ell+3)}}\langle\ell,1,\ell'+1\rangle_m\\
& \ -(\ell'+1)\sqrt{\frac{\ell'^2-m^2}{(2\ell'+1)(2\ell'+1)}}\langle\ell,1,\ell'-1\rangle_m \Bigg\}.
\end{aligned}
\end{align}
Again, following the notation of Ref.~\cite{Dolan:2018dqv}, we have defined $\langle\ell m | \hat{x}(\theta)|\ell' m\rangle:=\int_{-1}^1d\cos\theta K_{\ell'm}(\theta)\hat{x}(\theta)K^*_{\ell m}(\theta)$, as well as
\begin{align*}
\langle \ell_1,\ell_2,\ell_3\rangle_m:=(-1)^m\sqrt{\frac{(2\ell_1+1)(2\ell_2+1)(2\ell_3+1)}{4\pi}}\\
\times \begin{pmatrix}
\ell_1 & \ell_2 & \ell_3 \\ 
0 & 0 & 0
\end{pmatrix} 
\begin{pmatrix}
\ell_1 & \ell_2 & \ell_3 \\ 
-m & 0 & m
\end{pmatrix},
\end{align*}
in terms of the usual Wigner 3-j symbol.

In order to find non-trivial solutions of Eq.~\eqref{anglequation}, we require the
kernel of $M_{\ell\ell'}$ to be non-trivial. This is achieved by
imposing $\det M_{\ell \ell'}=0$, which restricts the separation constant $\nu$
in the complex plane. The connection of the Proca field solution to the
massless and longitudinal polarizations was established in
Ref.~\cite{Dolan:2018dqv}. The identification of a certain field mode is achieved in
the non-relativistic limit $\mu M\ll 1$ by matching the associated $\nu$ to the
following three cases. The parity-even $S=-1$ modes approach 
\begin{align}
\nu\rightarrow\nu_{-1}=\frac{- \omega}{m-a\omega} \ ,
\end{align}
in the non-relativistic limit for any $m=\ell$ mode. The parity-odd
polarization modes $S=0$ have 
\begin{align}
\nu_{0}=\frac{1}{2a}\left(m+1-a\omega+\sqrt{(a\omega-m-1)^2+4a\omega}\right),
\end{align}
when $\mu M\ll 1$. Finally, the quasi-bound non-relativistic limit of the $S=+1$ mode is given by the middle root $\nu_{+1}$ of the cubic \cite{Dolan:2018dqv}
\begin{align}
a \nu^3(m-a\omega)& -\nu^2((m+1)(m+2) \nonumber \\
& -a\omega(2m-a\omega))+\omega \nu+\omega^2=0.
\end{align}
These, together with Eq.~\eqref{freqguesses} for the frequency, serve as good
initial guesses in the non-relativistic limit for iteratively solving the
angular and radial Proca field equations.

Expanding the radial equation around the horizon at $r=r_+$ and infinity
$r\rightarrow\infty$ yields the asymptotic behavior and boundary conditions for
the radial solution.  These are given by
\begin{align}
R(r)\sim
\begin{cases} e^{-i \omega r_*}, & r\rightarrow r_+, \\
r^{(2\omega^2-\mu^2)M/Q}e^{-Qr}, & r\rightarrow \infty.\end{cases}
\end{align}
where $Q:=\sqrt{\mu^2-\omega^2}$, $r_*$ is the radial tortoise coordinate
\begin{align}
r_*(r):=r+\frac{2M r_+}{r_+-r_-}\ln\frac{r-r_+}{2M}-\frac{2Mr_-}{r_+-r_-}\ln \frac{r-r_-}{2M} \ ,
    \label{eq:tort}
\end{align}
and $r_{\pm}=M\pm\sqrt{M^2-a^2}$.  We utilize a direct integration method, and
impose the boundary condition at $r\rightarrow\infty$ by iteratively searching
for the complex value of $\omega$ that minimizes $\log |R(r_\text{max})|$ for
some large $r_\text{max}$.

\subsection{Proca field solutions} \label{procaresults}
For studying the astrophysical consequences of the superradiant instability, we
are most interested in the growth rate $\omega_I$ and real frequency $\omega_R$
of the Proca field solutions.  The former determines whether a given mode grows
sufficiently to liberate a non-negligible amount of the BH's energy, and be
potentially observable.  The latter dictates the frequency of the emitted
gravitational radiation since $T_{\mu\nu}\sim e^{2i\omega_R t}$ for a given
Proca field mode with frequency $\omega_R$ (if the cloud is dominated by a
single field mode).  Therefore, in the following, we emphasize the results for
the frequencies for several Proca masses, rather than focusing on the spatial
distribution or polarization states. 

\begin{figure}[t]
\includegraphics[width=0.48\textwidth]{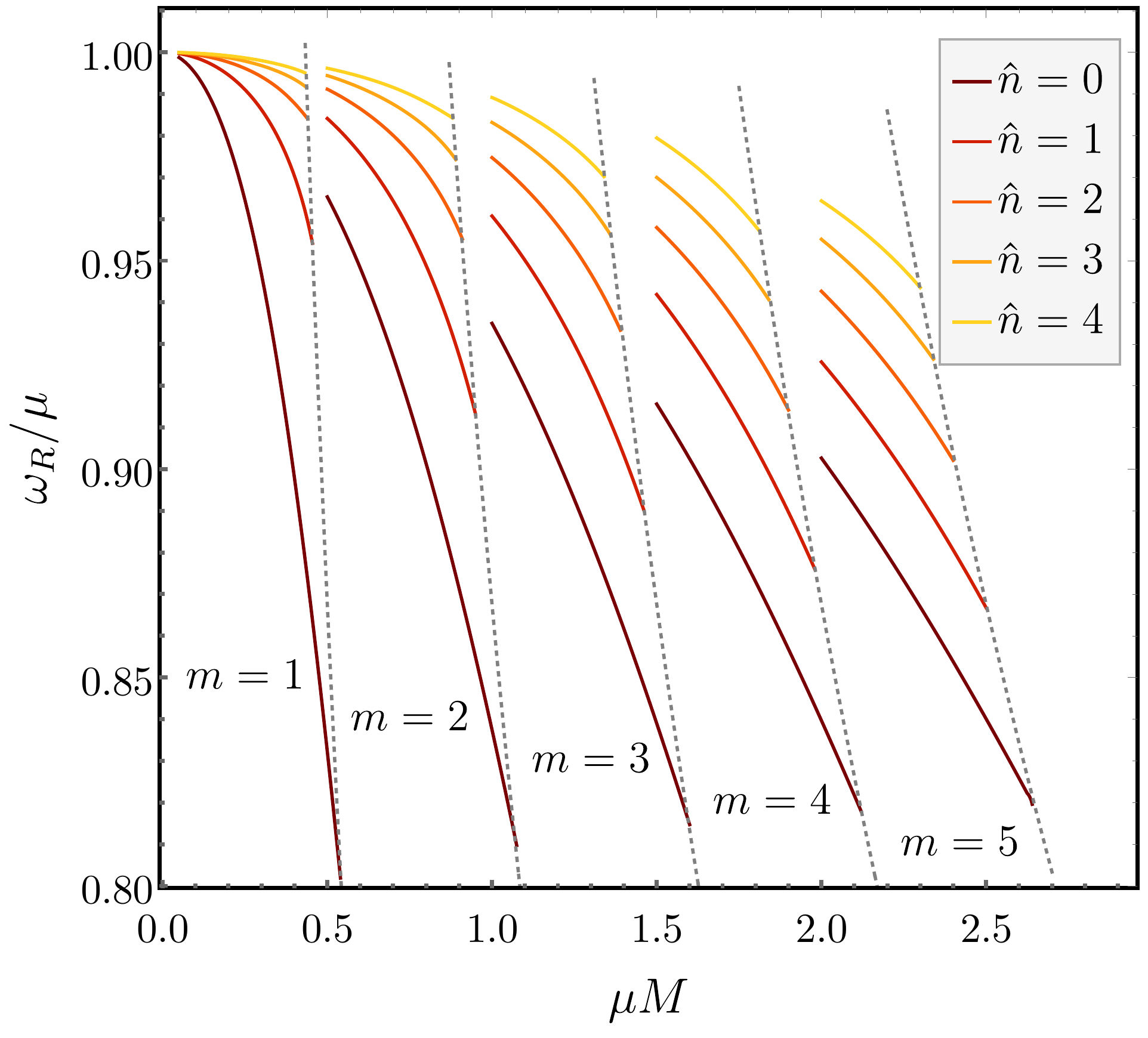}
\caption{
    The harmonic part $\omega_R$ of the Proca frequency, as a function of
    Proca mass parameter $\mu M$ on a Kerr background with $a=0.99M$. 
    For each $m$ mode, the first few overtones $\hat{n}$ are
    depicted for $S=-1$ (i.e., fastest growing) polarization state. The gray
    dashed lines indicate when the mode reaches saturation.
}
\label{realpart}
\end{figure}

The real parts of the frequencies of oscillation $\omega_R$ are presented in
\figname{\ref{realpart}}. As can be seen there, there exists a clear hierarchical 
structure of the $\omega_R$ among the overtones for a given $m$: 
The real frequencies of
the vector boson cloud are monotonically decrease with increasing overtone number
$\hat{n}$. This is to be expected in the non-relativistic limit, but it holds also
in the relativistic regime. 
This ensures that the superradiance condition $\omega_R< m\Omega_H$, for a
fixed $m$-field mode, restricts the higher overtones to saturate before the
lower ones.

\begin{figure}[b!]
\includegraphics[width=0.48\textwidth]{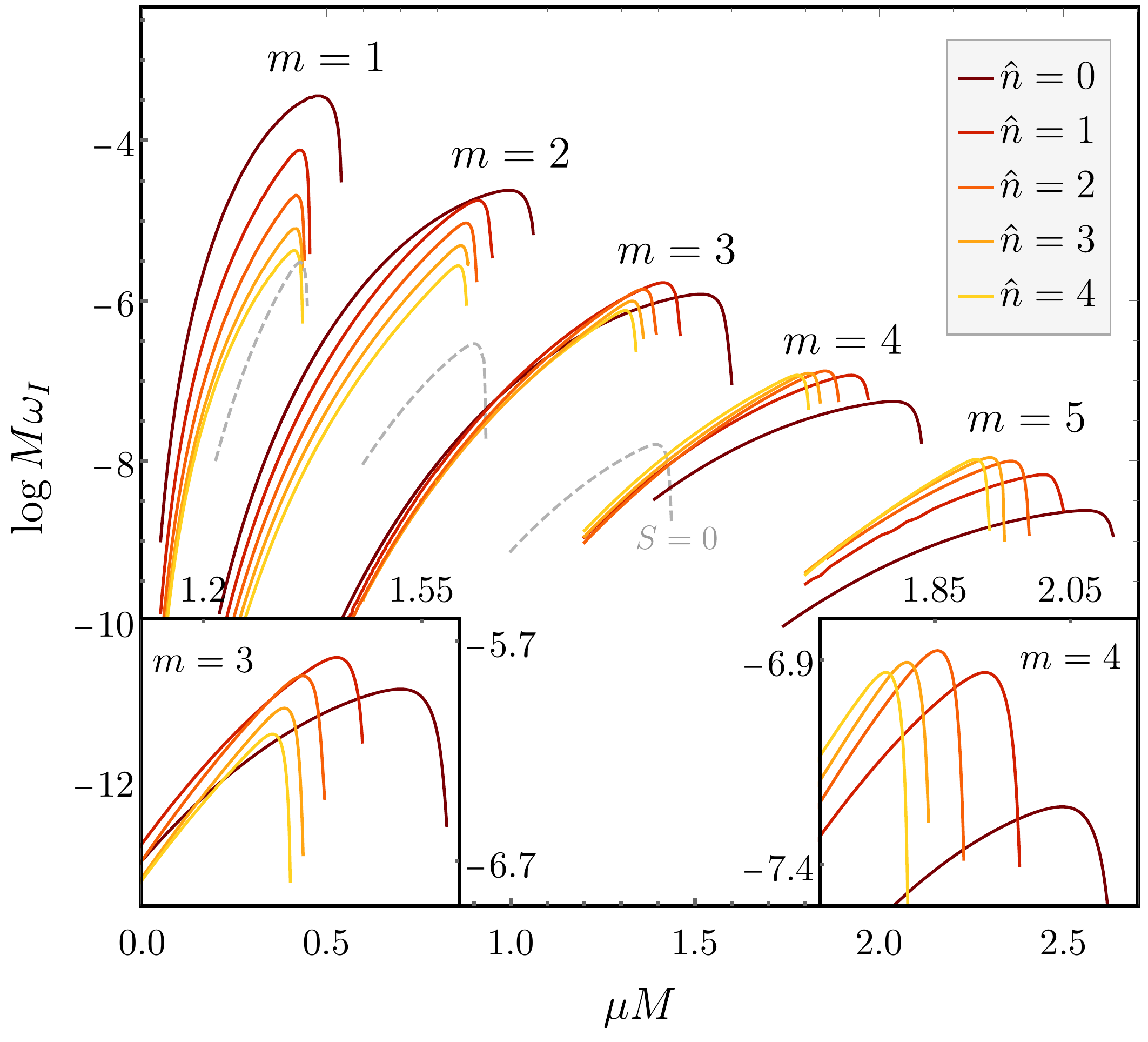}
\caption{
    The growth rates of the dominant polarization state, $S=-1$, of the
    cloud for each $m$-mode over a range of Proca masses $\mu$. We assume the
    Kerr parameter $a=0.99 M$, and include the first four overtones $\hat{n}$, as
    well as the fastest growing mode for the subleading polarization state
    $S=0$ (gray dashed line).
}
\label{growthrates}
\end{figure}

\begin{figure*}[t!]
\includegraphics[width=1\textwidth]{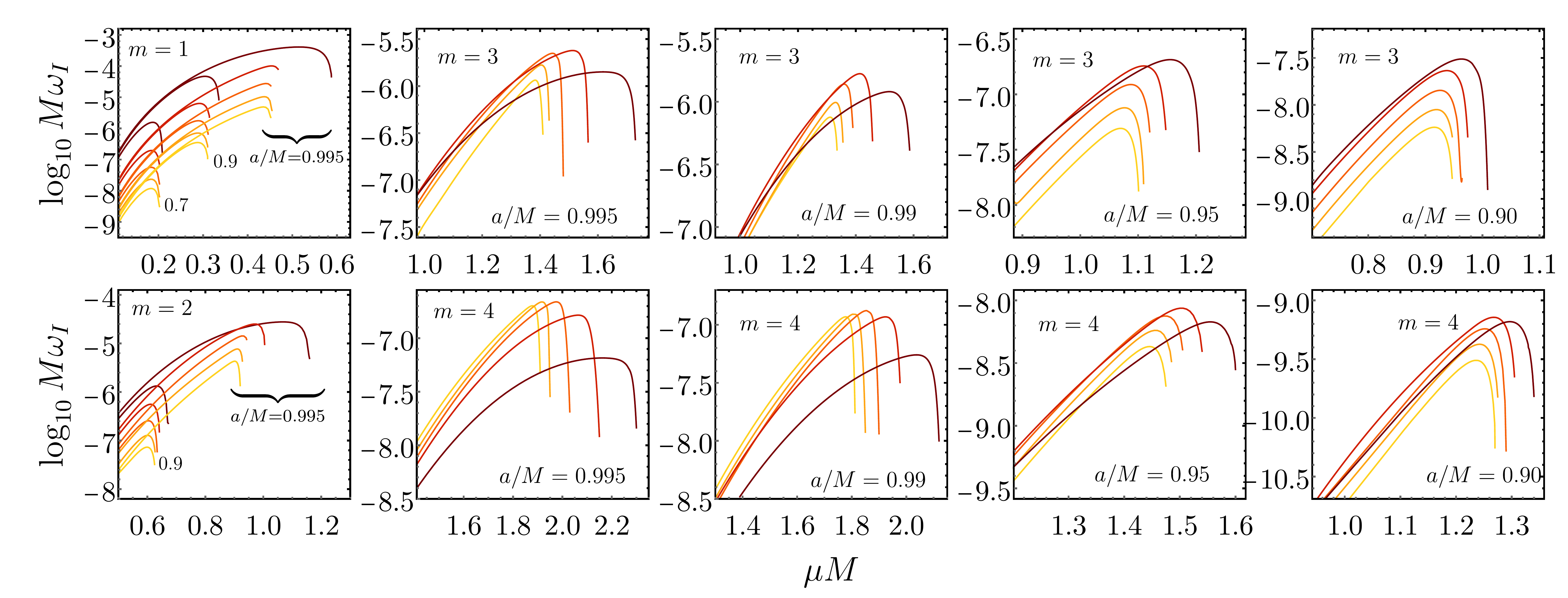}
\caption{
    The growth rates of the first five overtones ($\hat{n}=0$, \ldots, 4) of each $m$-field
    mode in the relevant Proca mass $\mu M$ range, and for several different
    dimensionless spin parameters $a/M$. For a description of the color-coding
    of the overtones see \figname{\ref{growthrates}}.
}
\label{modespins}
\end{figure*}

\figurename{ \ref{growthrates}} shows the instability growth rates for the
fundamental field mode in the $S=-1$ polarization state, as well as several
overtones, for different values of $m$. As already pointed out in 
Ref.~\cite{Dolan:2018dqv}, the $S\in\{0,+1\}$ polarization states grow more
slowly due to the fact that they are spatially
concentrated farther from the BH horizon (and therefore from the
source of energy and angular momentum). This example with $a=0.99M$ is typical 
of the behavior found
for lower and higher Kerr spin parameters. The first two azimuthal Proca modes,
$m=1$ and 2, have a strict hierarchical structure for this spin. The zeroth overtone, $\hat{n}=0$,
is the most dominant, whereas with increasing overtone number $\hat{n}$, the
instability growth rate decreases. Already in the $m=2$ case, the first signs
of a switching of the most dominant mode between the overtones is noticeable.
This effect of ``overtone mixing" becomes more pronounced with increasing
azimuthal number. In the $m=4$ case, for instance, the zeroth overtone is, in
fact, the slowest growing mode out of the first five overtones. The hierarchy
changes significantly with $\mu M$ and with spin parameter $a$. 
This overtone mixing behavior is captured by \figname{\ref{modespins}} for
several different BH spins. The hierarchical structure of the $m=1$ overtones
is preserved for all spin parameters $a\leq 0.995M$. However, already in the
$m=2$ case, the fundamental mode ($\hat{n}=0$) mixes with the first overtone
for BHs with near-extremal spin, with
$\omega_I^{\hat{n}=1}>\omega_I^{\hat{n}=0}$ for certain choices of $\mu M$.
For $m\geq 3$, the clear structure in the overtones is spoiled even for only
moderately high spins.  The fundamental mode tends to be subdominant at least
in parts of the relevant Proca mass parameter range. Importantly, however, the
$\hat{n}=0$ mode remains superradiant the furthest into the large $\mu
M$-regime. This ensures that it is the last mode to reach saturation of the
superradiance condition, and it can therefore grow significantly \textit{after}
the more dominant overtones have already reached saturation.  Again, this is
most pronounced around rapidly spinning BHs. It allows for the population of
\textit{several} different Proca field modes with significant amount of energy
and angular momentum, and ultimately leads to beating GW signals, as each such
populated Proca field mode has a slightly different real frequency $\omega_R$;
we discuss this further in Secs.~\ref{transitionmodes} and \ref{GWtransitions}.

Based on these data, we have constructed ready-to-use fit functions for the
real and imaginary parts of the frequency for the most dominant unstable
overtones for $m=1$, 2, and 3 (and $\hat{n}=0$). We employ a fitting function
similar to that used in Ref.~\cite{Cardoso:2018tly}, with a few modifications,
achieving more accurate fits to the growth rates. Details on the fits and their 
accuracy can be found in the Appendix. 

\subsection{Evolution of instability and normalization} \label{evolution}
The initial conditions for the evolution of the cloud from an initial seed field
configuration, to the macroscopic cloud, is, at the linear level, equivalent to
an energy distribution across the different available modes. This includes (for
a bound system) exponentially growing, as well as decaying, modes that could,
in principle, carry different angular momenta. The choice of the angular
momentum and mass distribution across the modes at the onset of the instability
affects the final state of the system (see Ref.~\cite{Ficarra:2018rfu} for an
analysis of multi-mode initial conditions for a scalar cloud). 
Here we assume that: (\textit{i}) the instability starts from a single vector
boson in each unstable mode, (\textit{ii}) the BH evolution is adiabatic and
can be described as a sequence of Kerr spacetimes, (\textit{iii}) the GW
emission to infinity and across the horizon is neglected during the growth of
the cloud, and finally, (\textit{iv}) any additional matter accretion of the BH
is ignored.
We justify these assumptions below.

The system evolves through $\sim 170$
e-folds, from a single
boson to the saturated cloud, which leads to an sensitive exponential dependence of the mass
per mode on the relative growth rates. 
Hence, as long as the initial seed field
(i.e., energy distribution) at the onset of the instability is small (e.g.,
condition (\textit{i}) is satisfied), the saturated state of the system is
expected to be completely dominated by a single, fastest growing, Proca field mode (see
Ref.~\cite{Ficarra:2018rfu}). An exception to this, which we discuss in the next sub-section, 
is when a mode with a lower instability rate can continue to grow after the saturation
of the fastest growing mode. Secondly, fully
non-linear evolutions of the set of Einstein-Proca equations revealed that the
adiabatic approximation (\textit{ii}) models the dynamics of the cloud's growth
even far from the test-field limit \cite{East:2018glu,East:2017ovw}. Furthermore, there is 
a clear separation of the GW emission and superradiant growth timescales 
$\tau_\text{GW}\gg\tau_\text{sup.}$. This implies (\textit{iii}), so GW
emission by the cloud (to infinity or through the BH horizon) can be ignored during the
exponential growth phase of the vector cloud. Finally, restriction (\textit{iv}) is merely a
simplification given the range of possible accretion rates and the fact that
we do not want to make any assumptions about the coupling of the massive vector boson
under consideration to the Standard Model.

As discussed in Sec.~\ref{introduction}, the ratio of the adiabatic change of a spinning
BH's mass $\delta M$ and angular momentum $\delta J$ due to the presence of a
\textit{single} field mode, with azimuthal number $m$ and frequency $\omega_R$,
is given by 
\begin{align}
\delta J/\delta M=m/\omega_R \ . 
    \label{changeofJandM}
\end{align}
The final BH mass $M_f=M_0+\delta M$ and angular momentum $J_f=J_0+\delta J$
at the saturation of the instability are therefore determined 
by solving for the value of $\delta M$ which gives the horizon frequency at which 
the superradiance shuts off:
\begin{align}
\Omega_H(M_f,J_f)=\Omega_H(M_0+\delta M,J_0+m\delta M/\omega_R)=\omega_R/m \ .
\label{saturatedstate}
\end{align}
In the above, $\omega_R$ is implicitly also a function of $M_f$ and $J_f$,
though the dependence on $J_f$ is small.
The explicit expression for the horizon frequency is 
\begin{align}
    \Omega_H(M,J)=\frac{J}{2M\left(M^2+\sqrt{M^4-J^2} \right)} \ .
\end{align}

In Eq.~\eqref{zansatz}, we indicated that we are solving for each Proca field
mode, characterized by the set of eigenvalues $\{\omega,\nu, m\}$, but we left the
overall normalization unspecified. A physical normalization of the modes is
achieved by choosing the value of $C^\nu_{m\omega}$ in Eq.~\eqref{zansatz} so
that the energy of the field solution matches the prediction of the above
established adiabatic approximation. The mass contained in the solution $A_\mu$
of the Proca field equation is, in Boyer-Lindquist coordinates, and using
Eq.~\eqref{energymomentumtensor}, given by
\begin{align}
E_A=-\int T^t{}_t(A) \sqrt{-g} drd\theta d\phi \ , 
\end{align}
where $g$ is the determinant of the Kerr metric. Then, for instance, the 
normalization at the time of saturation follows
immediately from requiring that $E_A=M_0-M_f$, assuming only a single
mode is populated at the saturation of the instability.

Beyond this normalization at the time of saturation, the temporal dependencies 
of the BH mass $M$ and angular momentum $J$, as well as the Proca cloud mass $E_A$, 
in the adiabatic context, are given by
\begin{align}
\dot{E}_A= & \ 2\omega_I(M,J)E_A,
\label{cloudexponential} \\
\dot{M} = & \  -\dot{E}_A, \qquad
\dot{J}=  m\omega^{-1}_R(M,J)\dot{M},
\label{bhmassexponential}
\end{align}
when considering a single Proca field mode. Therefore, once a single mode is 
normalized at the time of saturation (when Eq.~\eqref{saturatedstate} is satisfied), the 
entire time dependence of the cloud's mass is fixed by Eq.~\eqref{cloudexponential} to 
be primarily exponential in nature. The implicit dependence of $\omega$ on the BH 
mass and angular momentum only becomes important when the cloud acquires a 
significant amount of energy from the BH; this happens in the very last stages 
before saturation (due to Eq.~\eqref{bhmassexponential}), and is, as noted above in 
\textit{(ii)}, well-modeled by the adiabatic approximation. The evolution of a two-mode 
cloud is discussed in detail in the following subsection.

\subsection{Transitions between overtones} \label{transitionmodes}
So far, we have assumed that a \textit{single} mode 
dominates the Proca cloud formed through superradiance.
In the case of $m=1$, the lowest overtone ($\hat{n}=0$) is associated with
the largest growth rate $\omega_I^{\hat{n}=0}$ across parameter space,
which justifies this assumption. However, as pointed out in
Sec.~\ref{procaresults}, for $m\geq 2$, overtones $\hat{n}\geq 1$ can be faster
growing than the respective zeroth overtone $\hat{n}=0$ for specific choices of
$\mu M$ and BH spin $a$ (see also \figurename{ \ref{growthrates}}). Since,
as shown in \figname{\ref{realpart}}, the
overtone frequencies $\omega_R$ increase monotonically with $\hat{n}$, 
higher overtones saturate before the lower ones,
and leave spun-down BHs that are unstable against the subsequent growth of
lower overtones. The transition from the higher to the lower overtones allows
for \textit{several} superradiant Proca field modes to be populated
\textit{simultaneously}.

\begin{figure}[t]
\includegraphics[width=0.47\textwidth]{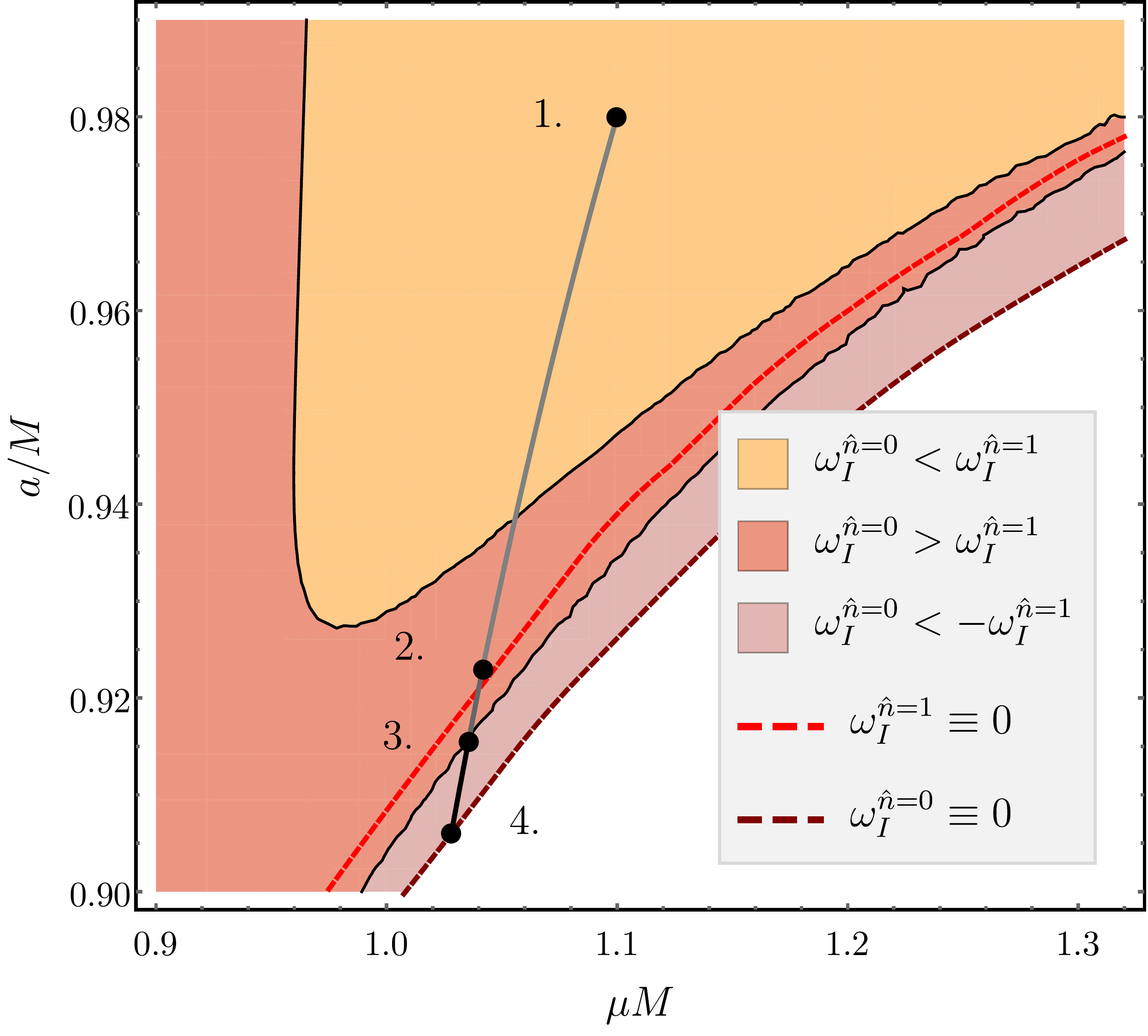}
\caption{
    An illustration of which overtone of the $m=3$ Proca field mode has the
    dominant growth rate as a function of dimensionless BH spin $a/M$ and mass
    parameter $\mu M$. The dark gray and black lines show the evolution in BH
    mass and spin of an example BH-Proca system, as described in the text. The
    red dashed line is the saturation point of the $\hat{n}=1$ overtone, while
    the brown dashed line is the saturation point of the $\hat{n}=0$
    overtone.
    \label{BHtranjectory}
    }
\end{figure}

To illustrate this, let us consider a BH with initial mass $M_0$ and
dimensionless spin $a_0/M_0=0.98$, as well as a Proca mass parameter $\mu
M_0=1.1$. From \figname{\ref{growthrates}}, we can see that in this regime the
$m=3$ field modes are the fastest growing, and that
$\omega^{\hat{n}=1}_I>\omega^{\hat{n}=0}_I$ for the given initial BH
parameters\footnote{Note, \figurename{ \ref{growthrates}} assumes $a/M=0.99$,
however, the qualitative behavior of the growth rates is the same for
$a_0/M_0=0.98$.}.  The initial BH state corresponds to point \textit{1.} in
\figname{\ref{BHtranjectory}}. Due to the fact that $\omega_I^{\hat{n}=0}>
\omega_I^{\hat{n}\geq 2}$ at point \textit{1.}, we consider the $\hat{n}=0$ and
$1$ modes only.  Hence, the Proca cloud's field solution is given by
$A^{\text{cloud}}_\mu =A^{\hat{n}=1}_\mu +A^{\hat{n}=0}_\mu$ at any given time
during the evolution of the instability.  The energy momentum distribution is
a sum of the energy momentum in $A^{\hat{n}=1}_\mu$ and $A^{\hat{n}=0}_\mu$, as
well as a cross term $T_{\mu\nu}^C$ defined by
\begin{align}
T_{\mu\nu}^C:=T_{\mu\nu}(A^1+A^0)-T_{\mu\nu}(A^1)-T_{\mu\nu}(A^0),
\label{energycrossterm}
\end{align}
with $T_{\mu\nu}(A)$ given by Eq.~\eqref{energymomentumtensor}, and where we
have relaxed the notation: $A^{\hat{n}=i}_\mu\rightarrow A^i_\mu$. Then, the
total mass of the two-mode cloud can be decomposed into
$E_A^\text{cloud}=E_A^{1}+E_A^{0}+E_A^C$. The time dependence of each of the 
individual components of $E_A^\text{cloud}$ is taken to be of the exponential form of 
Eq.~\eqref{cloudexponential}, with 
their associated growth rates $\omega_I$. We choose the initial conditions
$E_A^1=E_A^0=m_\gamma$ to be the mass of a single vector boson (though the same
behavior is expected from any sufficiently small initial field configuration).
Due to the spatial overlap of $A^1_\mu$ and $A^0_\mu$, we find that $E_A^C\sim
0.01 E_A^1$ at the onset of the instability. Starting with these initial data,
the BH moves from \textit{1.} to \textit{2.} in \figname{\ref{BHtranjectory}}
as the instability grows. 

\begin{figure}[t]
\includegraphics[width=0.48\textwidth]{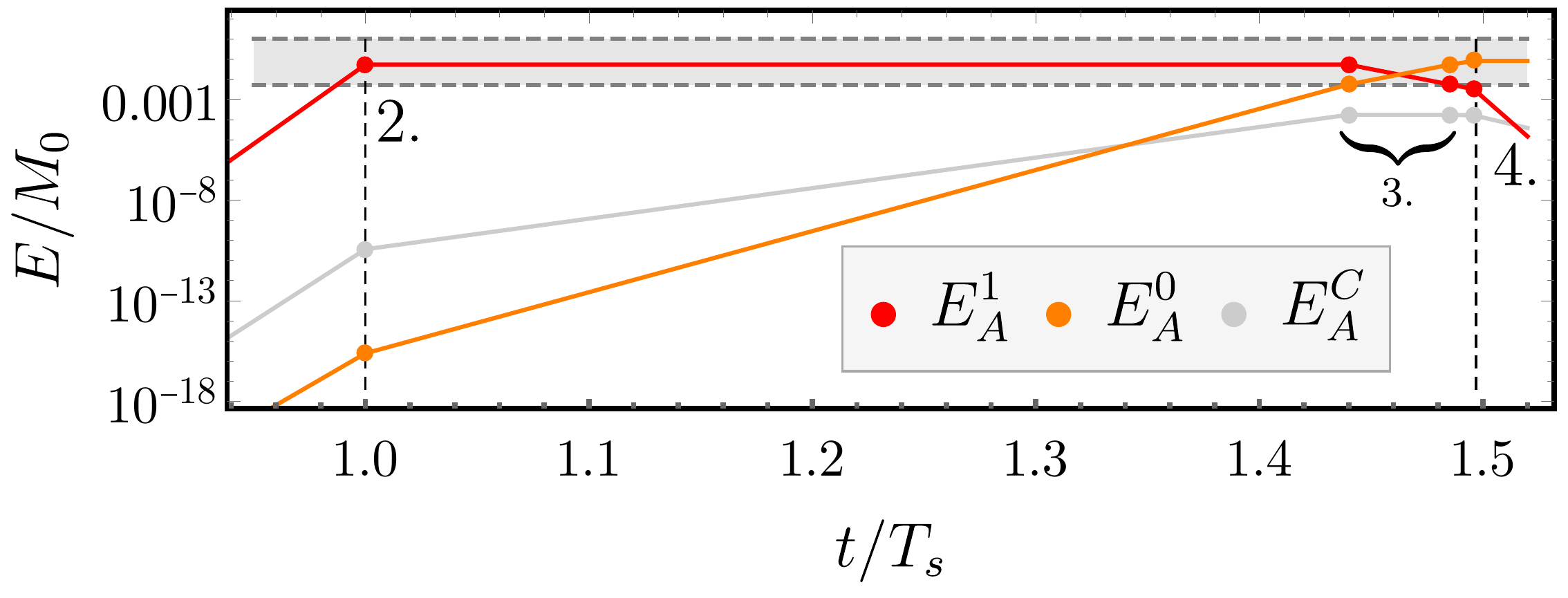}
\caption{
    The evolution of the components of the energy in a $m=3$ Proca cloud
    corresponding to the three terms in Eq.~\eqref{energycrossterm} for the example outlined in the text and \figname{\ref{BHtranjectory}}.
    Time $t$ is shown in terms of $T_s$ (the time it takes for the
    $\hat{n}=1$ overtone to saturate); in this case, 
    $T_s=0.51 (M_0/M_\odot)[1+0.012\log(M_0/M_\odot)]$ hours.
    The gray band, bounded by gray dashed
    lines, indicates the region from $E=5\times 10^{-3} M_0$ to $E=M_0$ (i.e.,
    the regime where the terms contain significant fractions of the initial BH
    mass $M_0$).}
\label{masstime}
\end{figure}

At point \textit{2.} in \figname{\ref{BHtranjectory}}, the $\hat{n}=1$ mode saturates with $E_A^1\approx
E_A^\text{cloud}=5.3\times 10^{-2} \ M_0$, where $E_A^0$ and $E_A^C$ are negligible, as shown in 
\figname{\ref{masstime}}. At this stage in the process,
$\omega_I^{\hat{n}=1}\approx 0$, while $\omega_I^{\hat{n}=0}>0$. Therefore,
$E_A^0$ and $E_A^C$ continue growing, while $E_A^1$ stays roughly constant\footnote{We
will see later that the GW time scale at this point is far lower than the
growth time scale of $\hat{n}=0$, so that $E_A^1$ remains constant in what
follows.}. As can be seen in \figname{\ref{masstime}}, $E_A^0$ and $E_A^C$ grow
until the BH approaches point \textit{3.} in \figname{\ref{BHtranjectory}}, 
where the \textit{decay} rate of the
$\hat{n}=1$ mode (since its frequency is now above the threshold for superradiance)
matches that of the \textit{growth} rate of $\hat{n}=0$:
$\omega^{\hat{n}=0}_I\approx|\omega_I^{\hat{n}=1}|$. Therefore, as the
$\hat{n}=1$ mode falls back into the BH, the $\hat{n}=0$ mode keeps extracting
energy and angular momentum from the BH. During this process, the spin and mass
of the BH stay roughly constant. This is because, on the one hand, if the BH
were to spin up due to the in-fall of $\hat{n}=1$, then the growth rate of
$\hat{n}=0$ would increase and counter the increase in BH spin. On the other
hand, if the BH is spun down significantly due to the growth of $\hat{n}=0$,
then the decay rate of $\hat{n}=1$ dominates, and the spin-down process is
damped. Therefore, the BH remains in the vicinity of point \textit{3.} throughout the entire
transition period, where all but a negligible amount of energy is transferred
from the $\hat{n}=1$ mode to the $\hat{n}=0$ overtone (see \figname{\ref{masstime}}). 
As a consequence, while the BH stays at point
\textit{3.}, both the $\hat{n}=1$ and the $\hat{n}=0$ overtones are populated
and carry significant fractions of the initial BH mass. This will lead to 
a a \textit{beating} GW signal which we discuss in detail in 
Sec.~\ref{GWtransitions}. Finally, once 
the mass in the $\hat{n}=1$ mode becomes negligible, the zeroth overtone
will spin down the BH further, until it reaches saturation at point \textit{4.}

Note that the results presented in \figname{\ref{masstime}} are based on assuming
the adiabatic approximation, and analyzing the different exponential
time-dependencies between points \textit{1.} through \textit{4.} Of course,
a non-linear evolution of the initial data described above is expected to show
a smoothing of the rough first approximation of the Proca cloud's dynamics
(see the non-linear study in Ref. \cite{Chesler:2018txn} for similar behavior
in the context of the anti-de Sitter superradiant instability).
Due to the exponential behavior, the dynamics of the instability is dominated only
by the modes containing a significant fraction of the BH mass, and therefore
the smoothing is expected to happen only very close to the individual points
\textit{2.} through \textit{4.} For instance, the $E_A^1$ term will hardly
decrease between points \textit{2.} and \textit{3.}, since $E_A^0$ is still
negligibly small, and only takes on significant values close to point \textit{3.}
Hence, the location and length of \textit{3.} in \figname{\ref{masstime}}, is
subject to the somewhat arbitrary cutoff of when a mode is considered to
contain a ``significant" amount of energy (we used a cutoff of $E=5\times
10^{-3} M_0$). However, other reasonable choices of this cutoff will not affect
the qualitative picture. 

These types of transitions from higher to lower overtones for a given azimuthal
mode number $m$ can be expected for all BHs with reasonably high spin and Proca
masses that fall into the regime where $m\geq 3$ modes dominate the
superradiance instability. This can be seen from \figname{\ref{growthrates}},
which shows the mixing of the growth rates of different overtones, with
$m\geq 3$, for the case of $a/M=0.99$. These can have astrophysical
significance as the growth rates even for some of the $m\geq 5$ azimuthal modes
are still well below the age of the Universe, even for supermassive BHs. 
We give a detailed analysis of the potential observability of these overtone
transitions in Sec.~\ref{GWtransitions}.

\section{Linear gravitational perturbations}

In the previous section, we solved the linear Proca field equations on a fixed
Kerr spacetime. The energy momentum tensor of the oscillating Proca cloud
resulting for superradiance will source GWs. In this section, we describe how 
we solve the linearized Einstein equations to obtain the amplitude, frequency, and
angular dependence of this gravitational radiation.

\subsection{Teukolsky formalism} \label{teukolskyformalism}
%
We compute the gravitational
radiation at future null infinity $\mathscr{J}^+$ using the Teukolsky formalism
\cite{Teukolsky:1973ha}, which describes linear perturbations of the Kerr background. The
Newman-Penrose (NP) Weyl scalar $\Psi_4:=C_{\alpha\beta\gamma\delta}n^\alpha
m^{*\beta} n^\gamma m^{*\delta}$, encoding the gravitational radiation at
$\mathscr{J}^+$, is projected out of the Weyl tensor
$C_{\alpha\beta\gamma\delta}$, with the Kinnersley null tetrad
\begin{align}
\begin{aligned}
n^\mu:= & \ \frac{1}{2\Sigma}(r^2+a^2,-\Delta,0,a),\\
m^\mu:= & \ \frac{\rho^*}{\sqrt{2}}(ia\sin\theta,0,1,i\sin^{-1}\theta),
\end{aligned}
\end{align}
where $\rho:=(r-i a \cos \theta)^{-1}$,
$\Sigma=r^2+a^2\cos^2\theta$, and $\Delta=r^2-2M r+a^2$ are the usual
Boyer-Lindquist functions. The Teukolsky equation for linear perturbations in
$\Psi_4$ decouple for $\Phi:=\rho^{-4}\Psi_4$ in terms of the eigenfunctions
$\{e^{-i\tilde{\omega}t},e^{i\tilde{m}\phi},{}_{-2}\tilde{S}_{\ell\tilde{m}}(a\tilde{\omega};\theta)\}$.
The $s=-2$ spin-weighted spheroidal harmonics\footnote{As in the angular
equation of the Proca field, the $\phi$-dependency is considered separately.}
are by definition solutions of the angular Teukolsky equation.  Here, and in the
following, we distinguish the eigenfunctions and eigenvalues of the Teukolsky
equations from the ones for the Proca field equations by a tilde. The
spin-weighted spheroidal harmonics are normalized to $2\pi$ over the 2-sphere
\begin{align}
\int^\pi_0|{}_{-2}\tilde{S}_{\ell \tilde{m}}(\theta)|^2\sin \theta d\theta=1.
\end{align}
In the Schwarzschild limit, i.e., setting the oblateness to zero,
$a\tilde{\omega}\rightarrow 0$, the eigenvalue ${}_{-2}A_{\ell \tilde{m}}$
corresponding to ${}_{-2}\tilde{S}_{\ell \tilde{m}}(\theta)$ (see 
Ref.\cite{Teukolsky:1973ha} for details) is given by ${}_{-2}A_{\ell
\tilde{m}}(a\tilde{\omega}\rightarrow 0)=\ell(\ell+1)+2$. Note that $\ell\geq
\max(|\tilde{m}|,2)$. The radial Teukolsky equation is explicitly given by
\begin{align}
\begin{aligned}
\Delta^2\frac{d}{dr}\left[\Delta^{-1}\frac{d\tilde{R}_{\ell\tilde{m}\tilde{\omega}}}{dr}\right]+V_{\ell\tilde{m}\tilde{\omega}}(r)\tilde{R}_{\ell\tilde{m}\tilde{\omega}}= \hat{T}_{\ell\tilde{m}\tilde{\omega}},\\
V_{\ell\tilde{m}\tilde{\omega}}(r)= \left[\frac{K^2+4i(r-M)K}{\Delta}+8i\tilde\omega r+{}_{-2}\lambda_{\ell\tilde{m}\tilde{\omega}}\right],
\label{radteueq}
\end{aligned}
\end{align}
where ${}_{-2}\lambda_{\ell\tilde{m}\tilde{\omega}}:={}_{-2}A_{\ell
\tilde{m}}+a^2\tilde\omega^2-2a\tilde{m}\tilde{\omega}$. The functions
$\hat{T}_{\ell\tilde{m}\tilde{\omega}}$ are the projection of the source
function $\tilde{T}$ for the Teukolsky master equation onto the basis of
eigenfunctions established above:
\begin{align}
\hat{T}_{\ell\tilde{m}\tilde{\omega}}=4\int_\mathbb{R} \frac{dt}{\sqrt{2\pi}}\int_{S^2} d\Omega {}_{-2}\tilde{S}^{*}_{\ell \tilde{m}}(a\tilde{\omega};\theta)e^{i\tilde{\omega}t-i\tilde{m}\phi}\frac{\tilde{T}(t,r,\Omega)}{\rho^{5}\rho^{*}}.
\end{align}
Note that the spheroidal harmonics are real in the case considered,
and $d\Omega:=\sin\theta d\theta d\phi$ is
the usual $S^2$ measure. Finally, the effective source $\tilde{T}(t,r,\Omega)$
in this NP description is computed from the projections of the stress-energy
tensor $T_\text{nn}:=T_{\mu\nu}n^\mu n^\nu,
T_{\bar{\text{m}}\text{n}}:=T_{\mu\nu}n^\mu \bar{m}^\nu$, and
$T_{\bar{\text{m}}\bar{\text{m}}}:=T_{\mu\nu}\bar{m}^\mu \bar{m}^\nu$ as
\begin{align*}
\tilde{T}(t,r,\Omega)= & -\frac{1}{2}\rho^8\bar{\rho}\hat{\mathcal{L}}_{-1}[\rho^{-4}\hat{\mathcal{L}}_0(\rho^{-2}\bar{\rho}^{-1}T_\text{nn})]\\
& -\frac{1}{2\sqrt{2}}\rho^8\bar{\rho}\Delta^2\hat{\mathcal{L}}_{-1}[\rho^{-4}\bar{\rho}^2\hat{\mathcal{J}}_+(\rho^{-2}\bar{\rho}^{-2}\Delta^{-1}T_{\bar{\text{m}}\text{n}})]\\
& - \frac{1}{4}\rho^8\bar{\rho}\Delta^2\hat{\mathcal{J}}_+[\rho^{-4}\hat{\mathcal{J}}_+(\rho^{-2}\bar{\rho}T_{\bar{\text{m}}\bar{\text{m}}})]\\
& - \frac{1}{2\sqrt{2}}\rho^8\bar{\rho}\Delta^2\hat{\mathcal{J}}_+[\rho^{-4}\bar{\rho}^2\Delta^{-1}\hat{\mathcal{L}}_{-1}(\rho^{-2}\bar{\rho}^{-2}T_{\bar{\text{m}}\text{n}})].
\end{align*}
The differential operators used here are
\begin{align}
\hat{\mathcal{L}}_s:= & \ \partial_\theta-i \sin^{-1}\theta \partial_\phi-ia\sin\theta \partial_t+s\cot \theta, \\
\hat{\mathcal{J}}_+:= & \ \partial_r-\Delta^{-1}[(r^2+a^2)\partial_t+a\partial_\phi] \ .
\end{align}

\subsection{Green's function} \label{greensfunctionsec}

As outlined above, the solution $\Phi$ to the Teukolsky equation can be
decomposed into the following eigenfunctions:
\begin{align}
\Psi_4=\rho^4\sum_{\ell,\tilde{m}\in \mathcal{K}} \int_\mathbb{R}\frac{d\tilde{\omega}'}{\sqrt{2\pi}}{}_{-2}\tilde{S}_{\ell \tilde{m}}(\theta)e^{-i \tilde{\omega}'t+i\tilde{m}\phi}\tilde{R}_{\ell \tilde{m} \tilde{\omega}'}(r),
\label{psi4}
\end{align}
where here, and in the following, we have defined 
$\sum_{\ell,\tilde{m}\in\mathcal{K}}:=\sum_{\tilde{m}=-\infty}^\infty\sum_{\ell=
\max(|\tilde{m}|,|s|)}^\infty$. 
This requires the inhomogeneous solution $\tilde{R}_{\ell
\tilde{m} \tilde{\omega}'}(r)$ of the radial Teukolsky
equation~\eqref{radteueq}. Since the separated Teukolsky equations are of
Sturm-Liouville type, we do this by constructing a Green's function from a set
of homogeneous solutions satisfying the considered boundary conditions.  As
usual, we focus on solutions that are purely ingoing at the horizon, and purely
outgoing at infinity. Therefore, following Ref.~\cite{Sasaki:2003xr}, we define
the two sets of homogeneous solutions 
$\{\tilde{R}^\text{in}_{\ell
\tilde{m}\tilde{\omega}}(r),\tilde{R}^\text{out}_{\ell
\tilde{m}\tilde{\omega}}(r)\}$ by the boundary conditions
\begin{align}
\tilde{R}^\text{in}_{\ell \tilde{m}\tilde{\omega}}(r)=
\begin{cases} B^\text{trans}_{\ell \tilde{m} \tilde{\omega}}\Delta^2e^{-ikr_*}, & r\rightarrow r_+ \\
B^\text{refl}_{\ell \tilde{m} \tilde{\omega}}r^3e^{i\tilde{\omega}r_*}+B^\text{inc}_{\ell \tilde{m} \tilde{\omega}}r^{-1}e^{-i\tilde{\omega}r_*}, & r\rightarrow \infty\end{cases}
\label{rinboundaryconditions}
\end{align}
and
\begin{align}
\tilde{R}^\text{out}_{\ell \tilde{m}\tilde{\omega}}(r)=
\begin{cases} C^\text{up}_{\ell \tilde{m} \tilde{\omega}}r^3e^{ik r_*}+C^\text{refl}_{\ell \tilde{m} \tilde{\omega}}\Delta^2 e^{-i k r_*}, & r\rightarrow r_+ \\
C^\text{trans}_{\ell \tilde{m} \tilde{\omega}}r^3 e^{i\tilde{\omega}r_*}, & r\rightarrow \infty\end{cases}
\label{routboundaryconditions}
\end{align}
with $k:=\tilde{\omega}-\tilde{m}\Omega_H$.  (Recall that $r_*$ is the radial
tortoise coordinate defined in Eq.~\ref{eq:tort}.) We obtain the homogeneous
solutions using the MST formalism
\cite{Mano:1996vt,Sasaki:2003xr,BHPToolkit}. In this context, we find that the
expansion of the homogeneous radial solution in terms of irregular confluent
hypergeometric functions around $r\rightarrow\infty$ exhibits faster
convergence behavior compared to the expansion around the horizon ($r=r_+$) in
the part of the parameters space considered here. This is true even relatively
close to the horizon. To check the validity of using the series expansion
around $r\rightarrow\infty$ even close to the horizon, we compare the resulting
solution with numerical integrations for the considered parameters and find
good agreement.

From Sturm-Liouville theory, the general solution, provided any well-behaved
source-function $\hat{T}_{\ell \tilde{m}\tilde{\omega}}$, is given by
\begin{align}
\begin{aligned}
\tilde{R}_{\ell \tilde{m}\tilde{\omega}}(r)=W_{\ell \tilde{m}\tilde{\omega}}^{-1}\Bigg\{ & \ \tilde{R}_{\ell \tilde{m}\tilde{\omega}}^\text{out}\int_{r_+}^rdr'\frac{\tilde{R}_{\ell \tilde{m}\tilde{\omega}}^\text{in}(r')\hat{T}_{\ell \tilde{m}\tilde{\omega}}(r')}{\Delta^2(r')}\\
+ & \ \tilde{R}_{\ell \tilde{m}\tilde{\omega}}^\text{in}\int_r^\infty dr' \frac{\tilde{R}_{\ell \tilde{m}\tilde{\omega}}^\text{out}(r')\hat{T}_{\ell \tilde{m}\tilde{\omega}}(r')}{\Delta^2(r')}\Bigg\},
\end{aligned}
\label{inhomogeneoussolution}
\end{align}
with the Wronskian $W_{\ell
\tilde{m}\tilde{\omega}}:=(\tilde{R}^{\text{in}}_{\ell
\tilde{m}\tilde{\omega}}\partial_r \tilde{R}^\text{out}_{\ell
\tilde{m}\tilde{\omega}}-\tilde{R}^\text{in}_{\ell
\tilde{m}\tilde{\omega}}\partial_r \tilde{R}^\text{out}_{\ell
\tilde{m}\tilde{\omega}})/\Delta$. By virtue of the radial Teukolsky equation,
$\partial_rW_{\ell \tilde{m}\tilde{\omega}}=0$.  Combining this with
the boundary conditions given by Eqs.~\eqref{rinboundaryconditions} and
\eqref{routboundaryconditions} implies that the Wronskian is $W_{\ell
\tilde{m}\tilde{\omega}}=2i\tilde{\omega}C_{\ell
\tilde{m}\tilde{\omega}}^\text{trans}B_{\ell
\tilde{m}\tilde{\omega}}^\text{inc}$. 
Therefore, the asymptotic solution at infinity is 
\begin{align}
\begin{aligned}
\tilde{R}_{\ell \tilde{m}\tilde{\omega}}(r\rightarrow \infty)& = \frac{r^3e^{i\tilde{\omega}r_*}}{2i\tilde{\omega}B^\text{inc}_{\ell \tilde{m}\tilde{\omega}}}\int_{r_+}^\infty dr' \frac{\hat{T}_{\ell \tilde{m}\tilde{\omega}}(r')\tilde{R}_{\ell \tilde{m}\tilde{\omega}}^\text{in}(r')}{\Delta^2(r')}\\
& =: \tilde{Z}^\infty_{\ell \tilde{m}\tilde{\omega}}r^3e^{i\tilde{\omega}r_*}.
\end{aligned}
\end{align}
While $\hat{T}_{\ell \tilde{m}\tilde{\omega}}$ has infinite support, the
integral is well-defined due to the exponential decay of the Proca field for
large $r$.  Because of the symmetry relation
$\tilde{Z}^\infty_{\ell,\tilde{m},\tilde{\omega}}=\tilde{Z}^{\infty
*}_{\ell,-\tilde{m},-\tilde{\omega}}$, we need only focus on either 
positive-$\tilde{m}$ or negative-$\tilde{m}$ solutions.

Once the amplitude of the asymptotic solution $\tilde{R}_{\ell
\tilde{m}\tilde{\omega}}(r\rightarrow \infty)$ to the inhomogeneous radial
Teukolsky equation is found, the gravitational radiation at future null
infinity $\mathscr{J}^+$ can be constructed. The Weyl-NP scalar at $\mathscr{J}^+$,
$\Psi_4^\infty:=\lim_{r\rightarrow\infty}\Psi_4$, behaves like
\begin{align}
\Psi_4^\infty(t,r,\theta,\phi)=\frac{1}{r}\sum_{\ell,\tilde{m}\in\mathcal{K}} \frac{\tilde{Z}^\infty_{\ell \tilde{m}\tilde{\omega} }}{\sqrt{2\pi}}e^{i \tilde{\omega}(r_*-t)+i\tilde{m}\phi}{}_{-2}\tilde{S}_{\ell \tilde{m}\tilde{\omega}}(\theta) \ ,
\label{psi4inf}
\end{align}
where we are focusing on a single frequency. 
We compute the GW radiation from the oscillations of different Proca modes (and
ignoring any imaginary frequency component). Since the stress-energy is
quadratic in the vector field, a single Proca mode with frequency $\omega_R$
and azimuthal number $m$ will source gravitational radiation with
$|\tilde{m}|=2m$ and $\tilde{\omega}=2\omega_R$.  (Hence $\tilde{\omega}$ can be
considered to be implicitly labeled by the mode index $\tilde{m}$ and overtone
number $\hat{n}$.)

\subsection{Cloud dissipation and gravitational radiation} \label{clouddissipation}

The evolution of the bosonic cloud can be split into two distinct phases,
characterized by two different timescales. First, the cloud superradiantly
grows in the unstable BH background with timescale
$\tau_\text{sup.}=1/\omega_I$; second, the mass and angular momentum dissipate
via GWs, characterized by the dissipation timescale $\tau_\text{GW}$. As
already argued above, the separation of scales,
$\tau_\text{GW}\gg\tau_\text{sup.}$, ensures that the GW dissipation can be
neglected throughout the entire first phase of the cloud's evolution.
Additionally, the gravitational wave energy flux through the horizon is
subdominant~\cite{Poisson:1994yf}, and hence
will be neglected in what follows.

With these assumptions, we can formulate the evolution of the bosonic
cloud during the dissipation process as: $\dot{E}_A=-\dot{E}_\text{GW}$. Here,
$E_A(t)$ is the energy contained in the Proca cloud,
whereas $\dot{E}_\text{GW}$ is
the average GW flux to future null infinity. In the Teukolsky formalism
described above, the GW energy flux averaged over several cloud oscillation periods is
given by
\begin{align}
\dot{E}_\text{GW}=\sum_{\ell,\tilde{m}\in\mathcal{K}} \frac{|\tilde{Z}^\infty_{\ell \tilde{m}\tilde{\omega}}|^2}{4\pi \tilde{\omega}^2}.
\label{averageflux}
\end{align}
Since $\dot{E}_\text{GW}\propto |Z^\infty_{\ell \tilde{m}\tilde{\omega}}|^2\propto E_A(t)^2$, 
integrating in time gives
\begin{align}
E_A(t)=\frac{E_A^\text{sat.}}{1+t/\tau_\text{GW}}, & &
	1/\tau_\text{GW}:=\dot{E}_\text{GW}(t=0)/E_A^\text{sat.},
\end{align}
where $E_A(t=0)=E_A^\text{sat.}$ is the saturation energy of the cloud and we
have defined the GW timescale $\tau_\text{GW}$ to be the time it takes for
half of the initial saturation energy of the cloud to be radiated away.

In performing this calculation, we also have to specify the mass of the
background geometry, i.e., the BH mass used when solving the Teukolsky
equation, which should lie somewhere in the interval $(M_f,M_0)$, where $M_f$ is
the BH mass at saturation.  Towards the end of the dissipation phase, i.e.,
when the bosonic cloud is mostly dissipated and therefore has a small mass,
the background BH has mass $M_f$ and angular momentum $J_f=a_f M_f$. The test
field approximation of the Proca field is applicable and the cloud's dynamics
are well-modeled by a background geometry with parameters $M_f$ and $a_f$. On
the other hand, the situation is not as clear at the onset of the dissipation.
At that stage the BH-Proca system has mass $M_0$ and angular momentum $J_0=M_0
a_0$, as the original BH. Hence, it can be expected that the energy contained
in the Proca cloud affects the background geometry.  This backreaction is not
properly accounted for in our test field approximation. We can, however,
approximate the \textit{combined} BH-Proca configuration by a background Kerr
geometry with unchanged mass and spin parameters $(M_0,a_0)$.  This is what we
do in the following, though we return to how the GW signal evolves with time in
Sec.~\ref{subsec:obs}.

\section{Results} \label{GWresults}
Using the methods outlined in the previous section for solving the linearized
Einstein equations, together with the Proca bound states obtained in
Sec.~\ref{procaresults}, we can compute the GW signals expected from ultralight
vector boson clouds.  In this section, we present results on the gravitational
radiation expected from clouds populated by a single mode or multiple modes, and
discuss the potential observability of such signals.

\subsection{Gravitational waves in the relativistic regime} \label{GWrela}

\begin{figure}[t]
\centering
\includegraphics[width=0.48\textwidth]{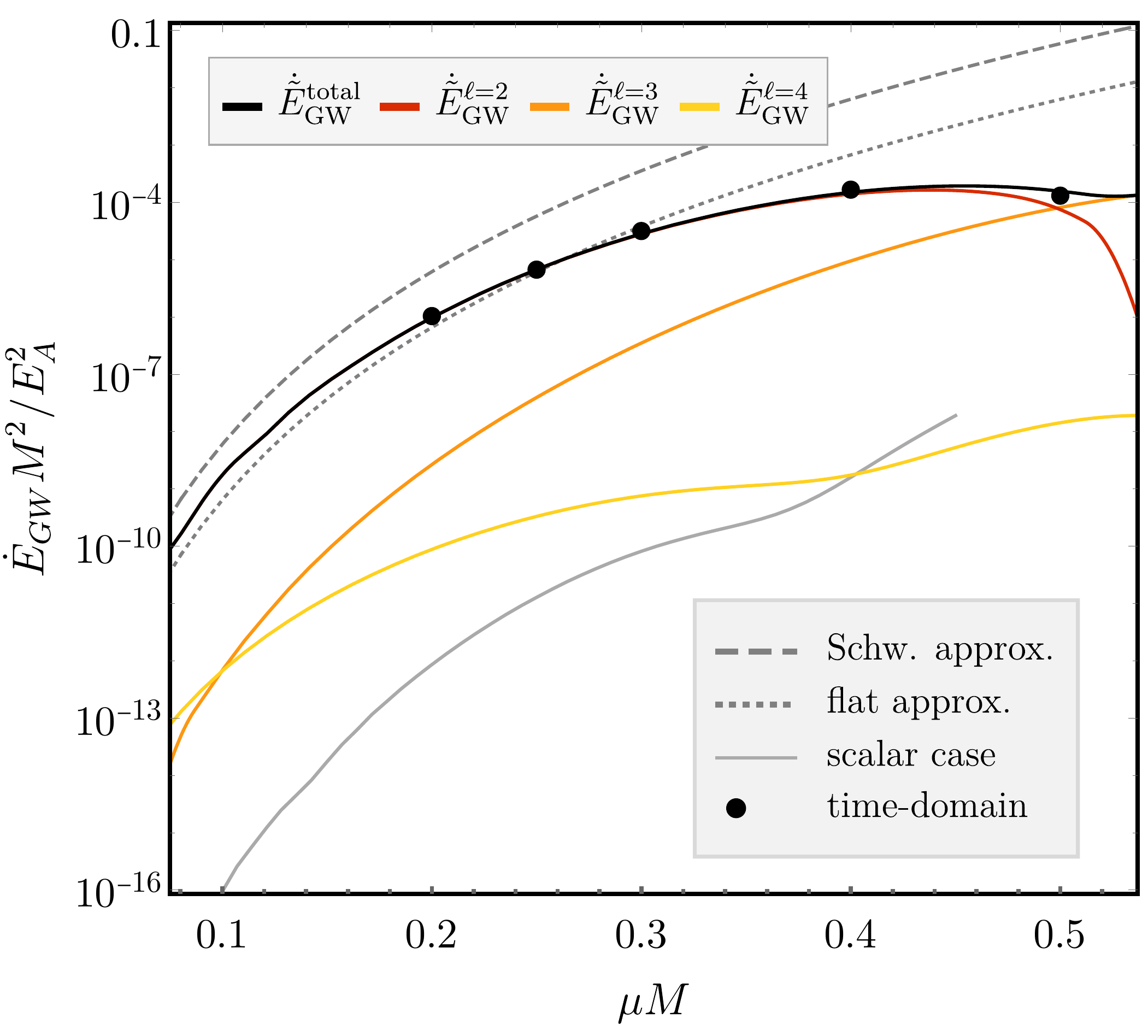}
\caption{
    The GW flux of the $m=1$ (and $\hat{n}=0$) Proca field mode around a Kerr
    BH with spin parameter $a/M=0.99$.  We show both the total GW
    flux $\dot{\tilde{E}}^\text{total}_\text{GW}$, and the individual contributions from
    spin-weighted spheroidal components with different angular numbers $\ell$
    (see Eq.~\eqref{averageflux}).
    For comparison we show the results from  time-domain simulations in the
	Proca-test field limit from Ref.~\cite{East:2017mrj} and the
	non-relativistic Schwarzschild and flatspace approximations from
	Ref.~\cite{Baryakhtar:2017ngi}.  Finally, we compare these vector boson
	results to the GW energy flux from a scalar boson cloud presented in 
	Refs.~\cite{Brito:2017zvb,Yoshino:2013ofa}.}
\label{flux99comparison}
\end{figure}

We begin by comparing the GW energy flux we obtain with previous numerical and
analytical results, as well as with the flux expected from an ultralight scalar
boson cloud. In \figname{\ref{flux99comparison}}, we show a comparison for the
case of a BH with spin $a/M= 0.99$ and the $m=1$ ($\hat{n}=0$) Proca field
mode. Note that we have normalized the GW flux by the cloud energy $E_A$ so as
to be independent of this quantity in the test-field limit: $\dot{\tilde{E}}_{GW}:=\dot{E}_{GW}M^2/E_A^2$.  Our relativistic
results match those obtained in Ref.~\cite{East:2017mrj} using time-domain
simulations, to within the estimated numerical error from the latter.
This also matches the full general relativity results for the GW luminosity from the Proca
cloud at saturation for $\mu M=0.4$ from Ref.~\cite{East:2018glu}.

In \figname{\ref{flux99comparison}}, we show two analytic approximations of the GW flux expected in the
non-relativistic limit from Ref.~\cite{Baryakhtar:2017ngi}.  Both are based on
the same bound Proca field states obtained in the long-wavelength regime, where
the massive wave equation in Kerr reduces to a set of Schr{\"o}dinger
equations. In the flat approximation, the flat d'Alembert Green's functions were
used to propagate the GWs from the source to null infinity. However, even the
leading order GW flux in the non-relativistic limit is affected by the curved
spacetime, so Ref.~\cite{Baryakhtar:2017ngi} also calculated an approximation
of the corrections from a Schwarzschild geometry (also shown).  The actual
GW flux in the non-relativistic limit is expected to fall in between these
values, which is consistent with our results.  In particular, as can be see
from \figurename{ \ref{flux99comparison}}, both the two small $\mu M$
approximations and our results follow the same powerlike behavior for $\mu M
\lesssim 0.15$.  However, when fitting $\beta \mu^{10} M^{10}$ to our numerical
results in $0.065 < \mu M <0.105$, we obtain a leading coefficient of
$\beta=16.66$, while $\beta_\text{flat}=32/5$ and $\beta_\text{Schw.}=60$ for
the flat and Schwarzschild approximation, respectively. Therefore, we further
improve on the non-relativistic limit here, by reducing the order-of-magnitude
uncertainty of the analytic results for $\mu M\ll 1$. Finally, for comparison,
we also include the corresponding GW energy flux for a scalar boson cloud
around a Kerr BH \cite{Yoshino:2013ofa,Brito:2017zvb}. Note that the
scalar field mode saturates at smaller $\mu M$, compared to the vector field
mode. Roughly speaking, the scalar GW power is six orders of magnitude smaller
than the vector GW power.

In our calculations of the GW energy flux, the dominant source of error emerges
from inaccuracies in calculating the Proca field solutions with the appropriate
boundary conditions. As discuss in Sec.~\ref{angularequation}, the boundary
condition for a bound Proca mode solution is enforced by minimizing the radial
solution evaluated at large $r$ over the complex frequency plane. The more
accurately the real and imaginary components of the frequency
$\omega=\omega_R+i\omega_I$ are known, the larger the accuracy of the Proca
field solutions. However, since we consider real frequencies with $\omega_R
M\gtrsim 10^{-2}$, but growth rates as low as $\omega_I M\sim
\mathcal{O}(10^{-10})$, the minimization over the complex $\omega$-plane is
limited by the precision of the floating-point calculation.  Therefore, we find
that in the non-relativistic limit, i.e., in the limit of small growth rate, the
Proca field solutions have larger relative uncertainties, which translates into
less accurate GW energy flux results in that limit. For instance, the total GW 
energy flux of
the $m=1$ mode, presented in \figname{\ref{flux99comparison}}, 
cannot be determined numerically with our approach for $\mu M<0.15$ without
making use of higher than double precision.  
As a result of the decreasing maximal growth rate
(with increasing azimuthal number $m$), the GW energy flux from higher-$m$
Proca field modes are less accurate.  We estimate the numerical error by 
varying the precision cutoff of the minimization algorithm for the Proca field 
solutions and observing the convergence of the resulting GW energy flux.  
Based on this, the relative uncertainty of the $m=1$ GW power is
$\lesssim 0.1 \%$ for $\mu M\in (0.4,0.6)$, $\lesssim 1 \%$ for $\mu M\in
(0.2,0.4)$, and $\lesssim 5 \%$ for $\mu M\in (0.065,0.2)$.

\begin{figure}[t]
\centering
\includegraphics[width=0.48\textwidth]{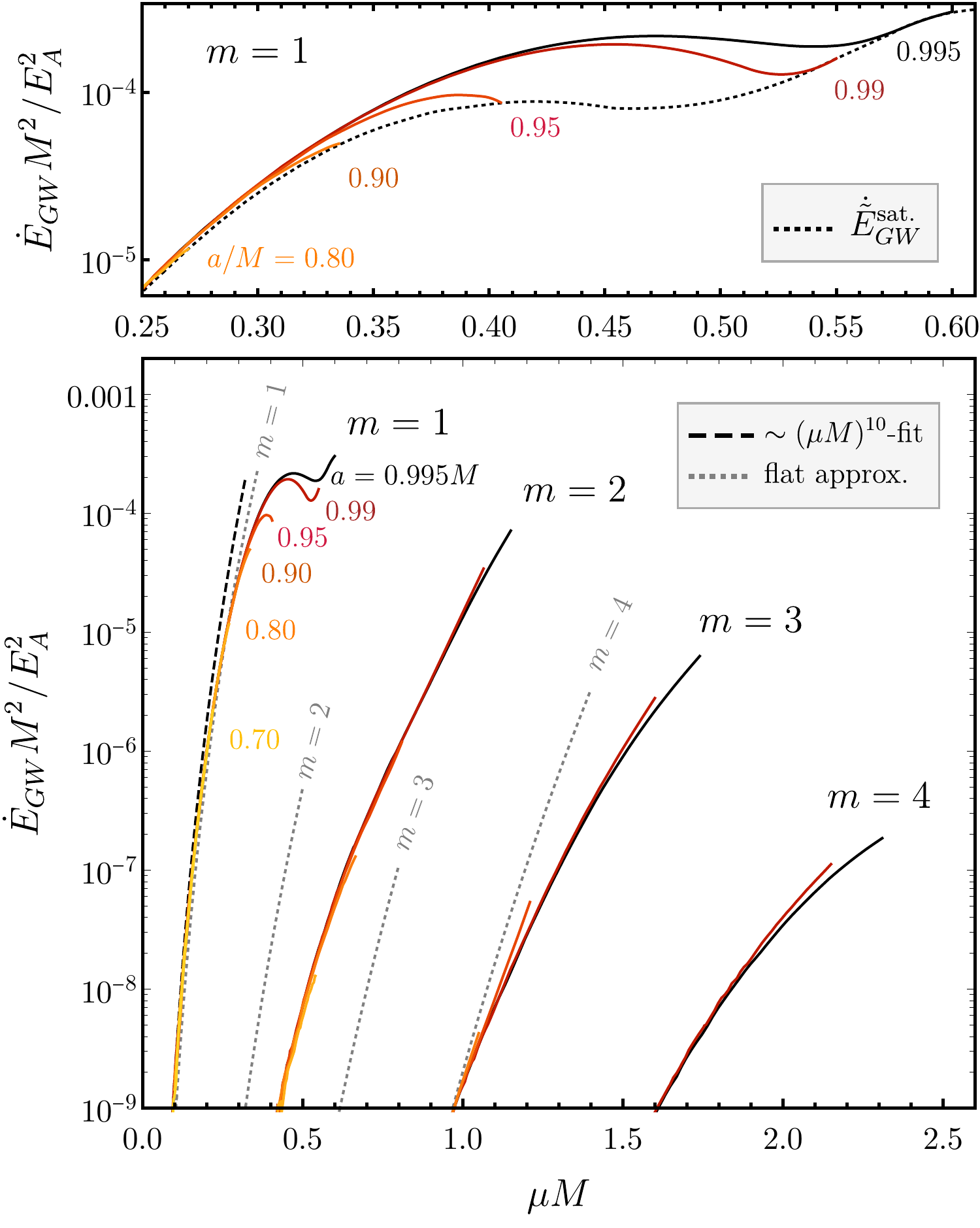}
\caption{
    The total rescaled GW energy flux $\dot{E}_\text{GW}\times(M/E_A)^2$
	emitted by a Proca cloud (solid lines) in the fundamental mode
	($\hat{n}=0$) for $m=1$, 2, 3, and 4.  We show results from BHs with
	spin $a/M\in\{0.7,0.8,0.9,0.95,0.99,0.995\}$. In the bottom panel, we also
	show the fit we obtained in the text and compare to the flat
	approximation of Ref.~\cite{Baryakhtar:2017ngi}. The
	relativistic features of the GW power from a Proca cloud in the $m=1$
	field mode are shown in more detail in the top panel, together with the
	mass-rescaled GW energy flux $\dot{\tilde{E}}^\text{sat.}_{GW}$ from a
	cloud that saturates the superradiant condition, i.e. with
	$\omega_R=\Omega_H$. 
    }
\label{overviewflux}
\end{figure}

In \figname{\ref{overviewflux}}, we present the GW power for different BH spins
and azimuthal Proca field modes. The energy flux is insensitive to the
dimensionless BH spin, except for near-extremal spins and large values of $\mu
M$, where some differences can be seen for the $m=1$ mode (though the GW power
from $m>1$ modes do not seem to share this behavior). In the top panel of
\figname{\ref{overviewflux}}, we also show the GW energy flux
$\dot{\tilde{E}}^\text{sat.}_{GW}$ when the cloud frequency matches the BH
rotational frequency $\omega_R=m\Omega_H$, as occurs at the saturation of the
superradiant instability.  This assumption sets the BH spin, and determines the
features of the GW power in the relativistic regime.
From these results, it is clear that the flat approximation
\cite{Baryakhtar:2017ngi} significantly overestimates the GW power in the
relativistic regime. Similarly to the $m=1$ case, we estimate the relative
uncertainties of the $m=2$ result
to be $\lesssim 1 \%$ for $\mu M\in (0.7,1.2)$ and $\lesssim 5 \%$ in the range
$\mu M\in (0.45,0.7)$. In the $m=3$ case, we estimate the uncertainty to be
$\lesssim 0.4\%$ for $\mu M\in (1.3,1.6)$ and $\lesssim 3\%$ for $\mu M\in
(0.9,1.3)$.\footnote{The $m=2$ results are less accurate, since the
$\ell=|\tilde{m}|$ spheroidal component dominates for those (as opposed to the
$\ell=|\tilde{m}|+1$ component in the $\tilde{m}=3$ and $\tilde{m}=4$ cases in
the considered $\mu M$ ranges), which seem to be more sensitive to
inaccuracies in the Proca field solution than the higher spheroidal components:
$\ell> |\tilde{m}|$.} Finally, the relative uncertainties of our GW power
results for a Proca cloud in the $m=4$ mode are $\lesssim 5 \%$ for $\mu M\in
(1.9,2.3)$ and $\lesssim 10 \%$ for $\mu M\in(1.6,1.9)$.

In addition, we calculate the GW power from vector boson clouds in excited states
for a BH spin of $a/M=0.99$ in \figname{\ref{overtoneflux}}. These indicate
that the clouds in excited states with larger growth rates also emit stronger
gravitational radiation compared to the fundamental modes. We also show the GW
power associated with the beating oscillations due to two modes being
simultaneously populated, which we discuss further in the next section.

\subsection{Gravitational waves from overtone transitions} \label{GWtransitions}

As we have seen in Sec.~\ref{transitionmodes}, there are parts of the BH-Proca
parameter space where several different overtones of a given azimuthal mode
number $m$ can be simultaneously populated. We continue the discussion of the example 
from Sec.~\ref{transitionmodes} here, and calculate the GW signal
produced during the transition of the cloud from overtone to fundamental mode. 

When the $\hat{n}=1$ overtone mode reaches saturation (point \textit{2.} in 
\figname{\ref{BHtranjectory}}), its energy is $E_A^1=5.3 \times 10^{-2} M_0$,
while the contributions from the fundamental mode are negligible.
Hence, the GWs are initially monochromatic with frequency given
by $2\omega^{\hat{n}=1}_R\big|_\textit{2.}=2.10/M_0$.
Since the GW emission timescale for
the $\hat{n}=1$ overtone is much larger than the growth timescale of the
$\hat{n}=0$ overtone (as already alluded to in
Sec.~\ref{transitionmodes}), the GW emission has no significant effect the cloud's dynamics. 

\begin{table*}[t]
\begin{ruledtabular}
\begin{tabular}{c|cc|ccccccccc}
	&$m_{\gamma}$ (eV)& $M_0$ ($M_{\odot}$)& $T_{\textit{1.}\rightarrow\textit{2.}}$ & $T_{\textit{2.}\rightarrow\textit{3.}}$ & $f^\text{GW}_{\textit{2.}\rightarrow\textit{3.}}$ (Hz) & $T_{\textit{3.}}$ & $f^\text{GW}_+$ (Hz) & $f^\text{GW}_-$ (Hz) & $T_{\textit{3.}\rightarrow\textit{4.}}$ & $f^\text{GW}_{\textit{3.}\rightarrow\textit{4.}}$ (Hz) \\ \hline
	LIGO & $1\times10^{-11}$& $92$           & 49.7 hours & $21.6$ hours & $731$ & $2.7$ hours & $725$ & $10.8$ & $36$ min. & $714$ \\ 
	LISA & $5\times10^{-17}$& $1.8\times 10^7$  & 1290 years & $567$ years & $3.66 \times 10^{-3}$ & $59$ years & $3.62 \times 10^{-3}$ & $5.38 \times 10^{-5}$ & $13.6$ years & $3.57 \times 10^{-3}$ \\
\end{tabular} 
\end{ruledtabular}
\caption{
	The duration and associated GW frequencies of the stages where
	different overtones are populated for the example discussed in the text
	with $\mu M_0=1.1$ and $a_0/M_0=0.98$.  We make two different choices
	for the physical mass of the vector boson $m_{\gamma}$ (from which the
	initial BH mass $M_0$ and all other quantities follow) corresponding to
	the LIGO and LISA bands.  The quantity
	$T_{\textit{1.}\rightarrow\textit{2.}}$ is the time for the BH to
	evolve from point \textit{1.} to \textit{2.} in
	\figname{\ref{BHtranjectory}} (i.e. for the saturation of the
	$\hat{n}=1$ mode), and similarly for the other times.  The different
	quantities $f_\text{GW}$ are the frequencies of the GWs emitted during
	the respective periods of time.  In particular, we denote the GW frequencies associated to
	$\omega^{\hat{n}=1}_R - \omega^{\hat{n}=0}_R$ and $\omega^{\hat{n}=1}_R
	+ \omega^{\hat{n}=0}_R$ with $f^\text{GW}_-$ and $f^{GW}_+$,
	respectively. 
	}
\label{beatingLIGOexamples}
\end{table*}

At $t\approx1.44 T_s$, where we recall that for this example 
$T_s=0.51 (M_0/M_\odot)[1+0.012\log(M_0/M_\odot)]$ hours,
the fundamental ($\hat{n}=0$) mode reaches one-tenth the energy of the
$\hat{n}=1$ mode (approaching point \textit{3.} in \figname{\ref{masstime}}).
From \figname{\ref{realpart}}, we see that the
harmonic frequencies of the overtones $\hat{n}=1$ and $\hat{n}=0$ are
different. Hence, the system, now consisting of two different field modes, emits
gravitational radiation at the four different frequencies
\begin{align}
\begin{aligned}
2\omega^{\hat{n}=1}_RM_0= & \ 2.10 ,\qquad & \ \omega_+M_0= 2.07 , \\
\omega_-M_0= & \ 0.03 , \qquad & \ 2\omega^{\hat{n}=0}_R M_0 =  2.04,
\end{aligned}
\label{fourfreqs}
\end{align}
where we have defined $\omega_+:=\omega^{\hat{n}=1}_R +\omega^{\hat{n}=0}_R$ and 
$\omega_-:=\omega^{\hat{n}=1}_R -\omega^{\hat{n}=0}_R$.
We illustrate how these frequencies and associated timescales would translate
into physical units in \tablename{ \ref{beatingLIGOexamples}}, for two
different choices of the vector boson mass that roughly correspond to the LIGO
and LISA bands, respectively. 

\begin{figure}[b]
\centering
\includegraphics[width=0.48\textwidth]{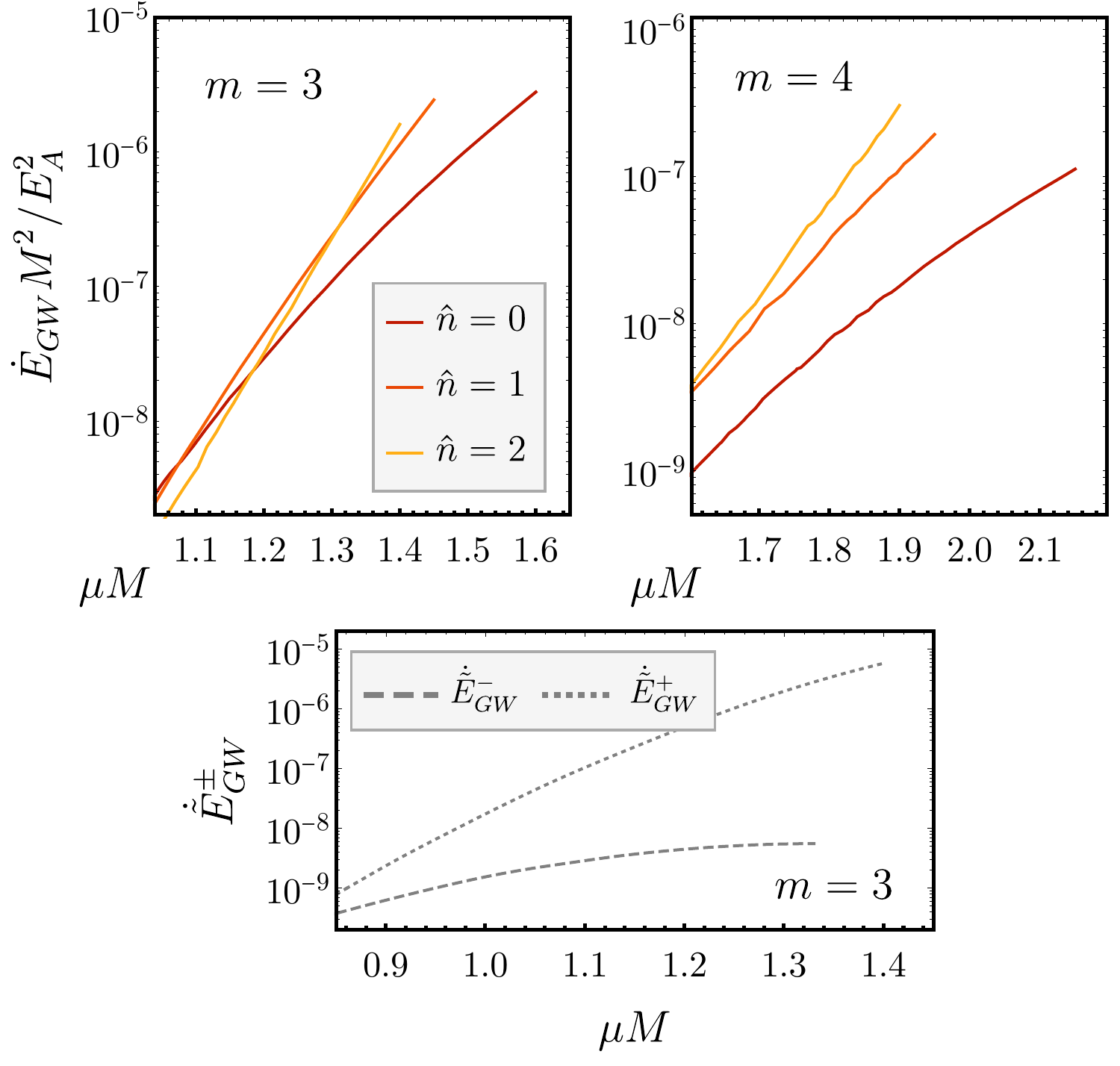}
\caption{
    The total rescaled GW energy flux $\dot{E}_\text{GW}\times(M/E_A)^2$
    emitted by a Proca cloud in the fundamental mode ($\hat{n}=0$) and the
    first two excited states, $\hat{n}=1$ and 2, for $m=3$ (top-left panel) and
    $m=4$ (top-right panel) and BH spin $a=0.99M$. In the bottom panel we show
    the mass-rescaled GW energy fluxes $\dot{\tilde{E}}^\pm_{GW}=\dot{E}^\pm _{GW}
    M^2/(E^1_A E^0_A)$ in the $\omega_\pm$ frequency components [as discussed
    below Eq.~\eqref{fourfreqs}] with BH spin $a=0.99M$.}
\label{overtoneflux}
\end{figure}

\figname{\ref{overtoneflux}} shows the mass-rescaled GW energy
fluxes, i.e., considering that $\dot{E}_\text{GW}^\pm\propto E_A^{1}E_A^{0}$,
in the four different frequency components for the special case of $a=0.99M$ as
a function of Proca mass. From this it is clear that the $\omega_+$ component
has the most normalized power, with
$\dot{\tilde{E}}^+_\text{GW}/\dot{\tilde{E}}_\text{GW}^i\sim\mathcal{O}(10^1)$
(where $i\in\{0,1\}$), while $\dot{\tilde{E}}^-_{GW}$ is the weakest of the
four.

\begin{figure}[b]
\centering
\includegraphics[width=0.48\textwidth]{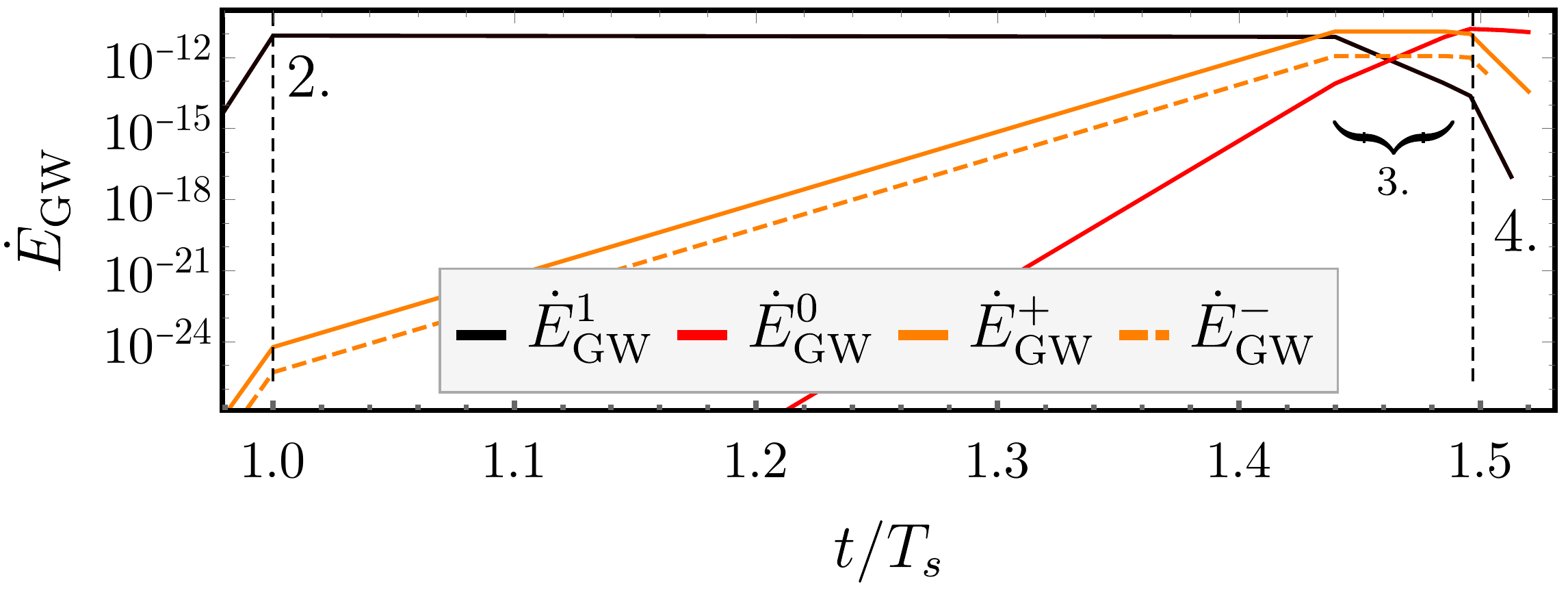}
\caption{
	The GW energy flux $\dot{E}_\text{GW}$ emitted during the overtone
	transition from $\hat{n}=1$ to $\hat{n}=0$ of the $m=3$ azimuthal Proca
	field mode for the example presented in Sec.~\ref{transitionmodes}.
	Here, $\dot{E}_\text{GW}^i$, with $i\in\{0,1\}$, corresponds to the
	frequencies $2\omega_R^{\hat{n}=i}$, while $\dot{E}^\pm_\text{GW}$ are
	associated with the frequencies $\omega_\pm$ (see Eq.~\eqref{fourfreqs}).
	}
\label{beatflux}
\end{figure}

Specifically for this example, we show the GW power $\dot{E}_\text{GW}$ in each of these frequency channels
before, during, and after the overtone transition in \figname{\ref{beatflux}}.
Here, we are taking into account the distribution of energy across the
two overtones (as opposed to scaling out the dependence on the cloud mass).
Note that, since we are using the simple approximation of the
evolution of the Proca cloud energy discussed in Sec.~\ref{transitionmodes},
the total power of the GW signal, illustrated in \figname{\ref{beatflux}}, does not change smoothly in time. However, as
discussed in Sec.~\ref{transitionmodes}, we expect that a fully non-linear
study of the mode transitions would reveal a smooth translation from the
single-frequency GW signal into the multi-frequency phase, and finally back to
the single-frequency GWs, until the cloud reaches saturation of the fundamental
mode. In
\figname{\ref{beatflux}}, we see that during the beating phase of the emitted
gravitational radiation (while the BH-Proca system remains at point
\textit{3.}), the frequency $\omega_+$ dominates the signal, while
$2\omega^{\hat{n}=1}_R$ slowly fades out and $2\omega^{\hat{n}=0}_R$ gains in
amplitude. The energy flux in the $\omega_-$ channel remains constant throughout the transition phase, just as the GW associated with $\omega_+$. However, the former is weaker than the latter, since $\dot{\tilde{E}}_{GW}^-\sim \dot{\tilde{E}}_{GW}^i$ in the relevant $\mu M$-range (see \figurename{ \ref{overtoneflux}}).
The overall GW power increase from \textit{2.} over \textit{3.} to
\textit{4.} This is due to the fact that, at their respective saturation
points, the $\hat{n}=0$ mode carries more energy compared with the $\hat{n}=1$
mode.

\subsection{GW frequency and drift}
\label{subsec:obs}

We now elaborate on the GW signals discussed in this section, and how they might
fall into frequency ranges observable by ground-based GW detectors like
LIGO for specific choices of boson mass and BH mass and spin.
(Our results can be similarly applied to future space-based GW detectors like LISA by considering
lower-mass bosons and supermassive BHs.)

\begin{figure}[t]
\centering
\includegraphics[width=0.48\textwidth]{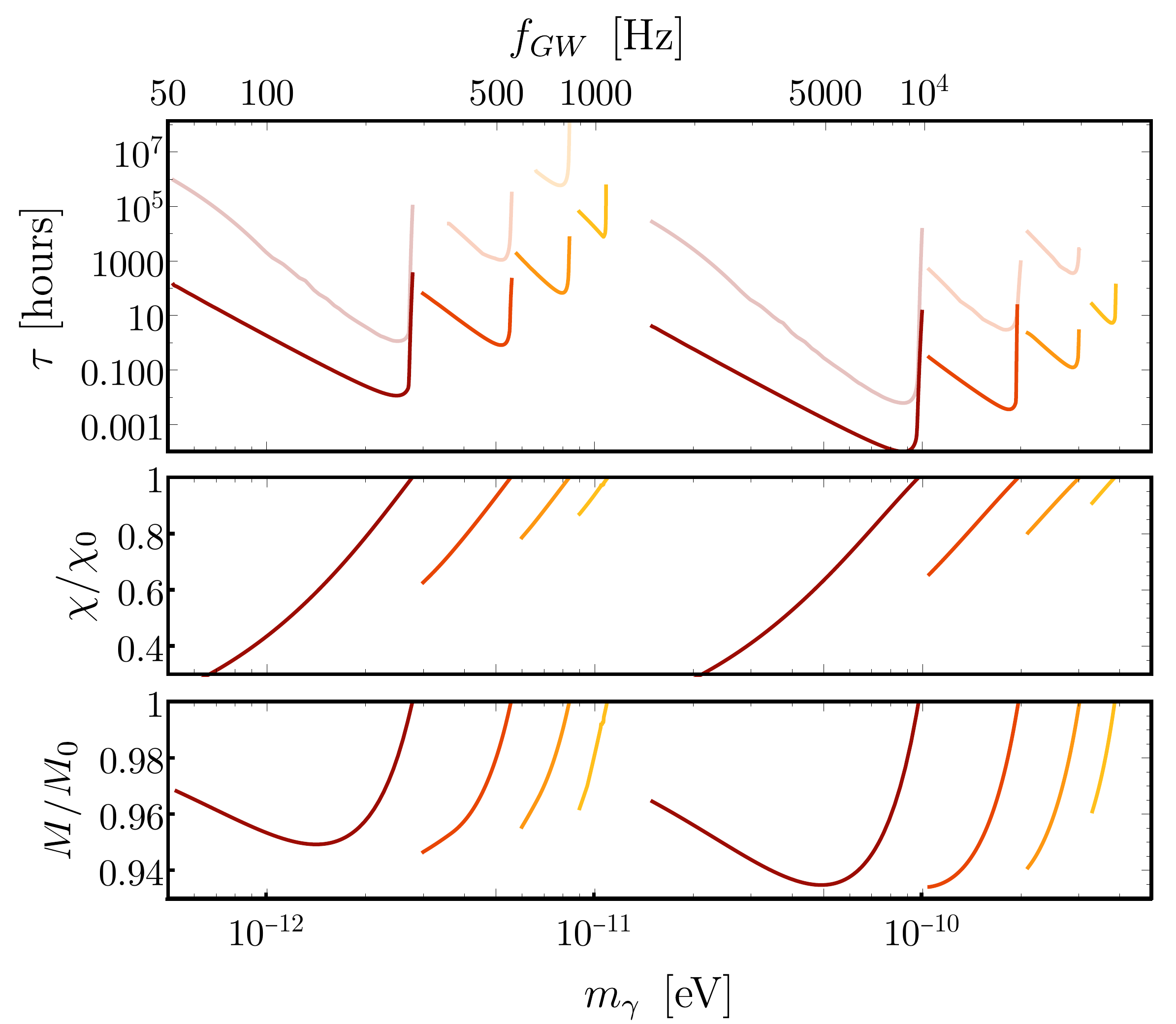}
\caption{
	Using the estimated values for the properties of the BHs produced from
	the merger events GW170729 (left) and GW170817 (right) as initial
	conditions for the superradiant instability, we present here the final
	BH parameters, $M$ and dimensionless spin $\chi$, in terms of initial
	mass $M_0$ and spin $\chi_0=J_0/M_0^2$ as a function of GW frequency
	$f_{\rm GW}$ and Proca mass $m_\gamma$. In the top panel, we show both
	the superradiance timescale and the GW timescale (where, for a given
	mode, the latter is the longer timescale) for each of the Proca field
	modes $m=1,\ \dots,\ 4$ ($m=1$ in dark red, up to $m=4$ in light
	yellow), with $\hat{n}=0$.  Note that, when constructing the top and
	bottom $x$-axis scales, we have approximated $\omega_R\approx \mu$.  We
	omit the GW timescales for radiation from Proca clouds in the $m=4$
	mode due to the larger numerical noise in the non-relativistic limit of
	our result (see Sec.~\ref{GWrela}).
	}
\label{intutition1}
\end{figure}

When a binary BH merger takes place, it creates a newly formed spinning BH
which, in the presence of an ultralight boson, is susceptible to the
superradiant instability, and could thus give rise to GWs from the resulting
boson cloud. In Fig.~\ref{intutition1}, we illustrate the expected properties
for two LIGO merger events on the high and low end of the remnant mass. The first
is GW170729, which gave rise to a BH with an estimated mass of $M_0=80\
M_{\odot}$ and dimensionless spin $\chi_0=0.81$~\cite{LIGOScientific:2018mvr}.
The second is the neutron star merger GW170817, which we assume gave rise to a
BH with $M_0=2.8\ M_{\odot}$ and $\chi_0=0.89$ (the upper end of the allowed
range)~\cite{LIGOScientific:2018mvr}, and ignore any possible effects from
matter remaining outside the BH, for the purpose of this illustration. 
For the larger remnant mass case (GW170729), we can see in particular that the
GWs fall into the range of LIGO's sensitivity (roughly 10 Hz to a few kHz) for
a sizable range of possible vector masses.  In many cases in this range, a few
percent of the BH's total mass will be converted into a boson cloud, and
ultimately radiated as GWs, on timescales that can be as short as hours. For
the small remnant mass (GW170817), larger putative boson masses are probed, 
resulting in timescales that are even shorter (as short as
minutes) and GW frequencies that are higher, typically above a kHz, where LIGO
is less sensitive.  Other remnant BHs with masses in between $3 M_{\odot}$ and $80\
M_{\odot}$ will have properties intermediate to these two cases.

So far we have assumed that the GWs from boson clouds (at least those dominated
by a single Proca mode) have constant frequency. However, there will be a small
increase in the GW frequency as the mass of the BH-Proca cloud system
decreases, due to the conversion of the cloud into GWs.  Roughly speaking, the
negative contribution to a boson's energy, and hence frequency, due to the
self-gravity of the cloud decreases, leading to higher frequency gravitational
radiation.  This frequency drift can be important for continuous GW searches,
and can limit the range of accessible sources when not taken into account,
especially for GWs from massive vector clouds (as opposed to massive scalar
clouds), due to their small GW timescales (and therefore, larger frequency
derivatives). We note that this positive frequency derivative of the (almost)
monochromatic GW signal stands in contrast to the negative frequency derivative
expected, for example, in a pulsar spinning down due to GW emission from a
mountain.

\begin{figure}[t]
\centering
\includegraphics[width=0.48\textwidth]{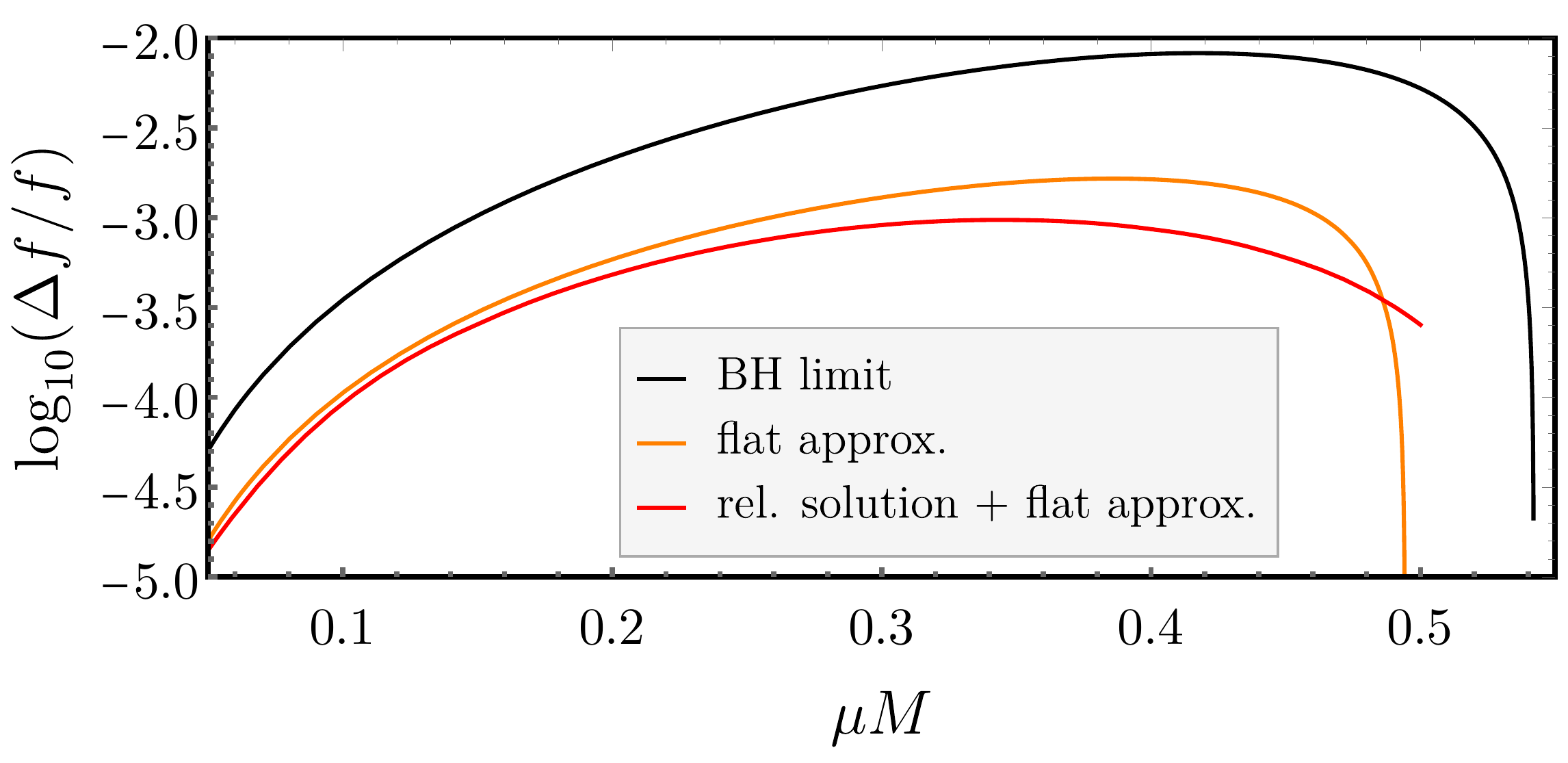}
\caption{
    The total relative frequency drift $\Delta f/f$ a monochromatic GW signal
    experiences during the first GW timescale $\tau_\text{GW}$ after
    saturation of the superradiant instability of the $m=1$ Proca field mode
    around a BH with $a/M=0.99$. In our approach of estimating $\Delta f$, labeled 
    ``BH limit", the spatial distribution of the cloud's self-energy is ignored,
    instead, it is added to the BH mass. For comparison, we also show the
    non-relativistic results obtained in Ref.~\cite{Baryakhtar:2017ngi} (labeled 
    ``flat approx."), as
    well as a similar calculation using the same Newtonian expression for the
    cloud's self gravity, but our relativistic expressions for the energy
    density and saturation energy (red line).
}
\label{freqdrift}
\end{figure}

As a first estimate of the expected change in GW frequency during the evolution
and eventual dissipation of a Proca cloud, we compute the difference in
frequency $\Delta f$ between the GWs obtained when solving on a BH background with the
original mass $M_0$ and spin $a_0$, and those obtained on a BH background with
mass decreased by half the saturation energy of the boson cloud
$M=M_0-E_A^\text{sat.}/2$ and an equivalently lower spin $a$ (as obtained
through Eq.~\eqref{changeofJandM}). This is shown in \figname{\ref{freqdrift}} 
(labeled as ``BH limit"). Note that
the real part of the frequency is hardly dependent on the dimensionless
background spin, making the decrease in mass of BH-Proca system the leading
contribution to the frequency drift.  This BH perturbation theory calculation
of course ignores the fact that during the GW emission phase, some of the mass
of the system is spread over the spatial extent of the cloud (as opposed to all
being in the BH), and thus overestimates the magnitude of the gravitational self-energy of the
cloud. In Ref.~\cite{Baryakhtar:2017ngi}, the frequency drift was estimated
in the non-relativistic limit using the Newtonian expression for the gravitational
self-energy of the cloud. For comparison, we also show their result in \figname{\ref{freqdrift}},
as well as the estimate we obtain using our relativistic expression for the cloud energy
density and cloud saturation energy $E_A$ in a similar calculation.

In \figname{\ref{freqdrift}}, we consider a BH-Proca system with $a/M=0.99$ in
the $m=1$ fundamental mode, though similarly large frequency changes occur also
for the zeroth overtones of $m=2$, 3, and 4.  In the non-relativistic limit,
the BH limit calculation is larger by a factor of $\approx 3$ compared to the
other calculations, just due to the fact that all the mass is assumed to be in
the BH, which is only valid at early times, before the boson cloud as obtained
significant mass, and at late times, after the boson cloud has mostly
dissipated.  For the other curves in \figname{\ref{freqdrift}}, which make use
of the Newtonian expression for the cloud's self gravity, replacing the flat
approximation of Ref.~\cite{Baryakhtar:2017ngi} by our relativistic solutions
of the Proca field gives a small difference (mainly due to the different values
of $E_A^\text{sat.}$), and actually decreases the estimated frequency change 
(except at large values of $\mu M$).

\section{Discussion and Conclusion}
\label{sec:discuss}
We have investigated the superradiant instability of a massive vector field
around a spinning BH with techniques from BH perturbation theory, and
utilizing an adiabatic approximation to model the evolution of the BH and Proca
cloud. We calculated the properties of the GWs that would be emitted
by vector boson clouds that arise through superradiance, including their frequency, 
amplitude, and evolution.

We found that the quasi-normal bound states of a Proca field around a Kerr BH
exhibit a rich overtone structure in the relativistic regime.  Excited states
($\hat{n}>0$) of the Proca field can, depending on the Proca mass and BH spin,
have the largest instability rate, and thus dominate the initial dynamics.
Since lower-$\hat{n}$ states always have lower frequency, and thus can continue
to grow after the saturation of higher frequency instabilities, this leads to
overtone mixing.  For $m=1$, the fundamental mode is the most unstable.
However, for $m=2$, this overtone mixing occurs only for near-extremal BH spin,
while for $m=3$, it occurs for moderately high BH spins.  For $m\geq4$ the
overtone mixing occurs even in the non-relativistic limit, and would produce
several overtone transitions.
Similar overtone mixing in the relativistic regime was found in the scalar
boson case for $m=3$ modes~\cite{Yoshino:2015nsa}. 

After the saturation of the fastest growing overtone, the Proca cloud undergoes
transitions between the excited states until the instability finally saturates
in the fundamental mode. Several modes can be populated simultaneously during
these overtone transitions, which, leads to a distinctive multi-frequency GW
signature.  As we have seen, the gravitational radiation in the transition
phases is just as strong as during the monochromatic phases, and the
frequencies could also fall into the LIGO/LISA bands. In fact, the beat
frequency can be orders of magnitude smaller than the individual overtone
frequencies, making signals from BH-Proca systems with heavier massive vector
bosons fall into a detectable frequency range when the individual frequencies
are too high frequency to detect.  However, the amount of time spent in this phase 
(and thus total radiated energy) is small, making observing them more difficult.  
For the specific example discussed in Sec.~\ref{GWtransitions}, 
the transition phase lasts only $4\%$ of the time
from the onset of the instability to the saturation of the zeroth overtone, 
and only $\approx 10^{-3} \tau_\text{GW}$ (the GW dissipation timescale of the 
cloud). 

In this work, we calculated the power of the GW emission from Proca clouds
around spinning BHs using the linearized Einstein equations, improving upon the
flat and Schwarzschild approximations from Ref.~\cite{Baryakhtar:2017ngi} in
the non-relativistic limit.  We also substantially expanded on the parameter
space covered compared to the previous relativistic calculations in
Refs.~\cite{East:2017mrj,East:2018glu}, studying a range of BH spins, different
Proca field modes (azimuthal numbers $m=1$, 2, 3, and 4) and overtones
($\hat{n}\geq 0$). We find that the GW energy flux normalized by the cloud mass
is mostly independent of the BH's spin. Our results also show that the flat
approximation significantly overestimates the energy flux from Proca clouds for
larger values of $\mu M$ (and this appears to get worse for increase values of
$m$).  For $m=1$, and small (non-relativistic) values of $\mu$, the flat 
approximation underestimates the GW power by a factor of a few.

Our results can be used in searches for the GW signals from ultralight vector
boson clouds around spinning BHs, which could fall into several different
categories.  First, targeted searches can be used to follow-up detections of
compact binary mergers, where the remnant is a spinning
BH~\cite{Ghosh:2018gaw,Isi:2018pzk}.  Using the measured properties of the BH,
these results can be used to predict the possible signals for different
putative boson masses.  In such targeted searches, the above described overtone
transitions, and associated distinctive GW signatures, should also be taken
into account.  In addition, all-sky continuous GW
searches~\cite{Goncharov:2018ufi,DAntonio:2018sff}, tailored to these sources,
could provide more stringent bounds on vector boson masses, similar to the
analysis in Ref.~\cite{Palomba:2019vxe} for the scalar case.  Finally, one can
search for, or place constraints on, ultralight vector bosons by looking for a
stochastic GW background, similarly to the scalar
case~\cite{Brito:2017wnc,Brito:2017zvb}.

Compared to ultralight scalars, vectors have a superradiance instability growth
rate and GW amplitude that is several orders of magnitude larger.  As we have
seen here, the timescale for a vector boson cloud around a stellar mass BH to
be converted into GWs can be as short as hours. Though the GW signal is louder,
the shorter duration, as well as the faster frequency increase in the signal
may be a challenge for conventional continuous GW searches, which typically
assume a signal that lasts the duration of an observing run.  This may be a
motivation for designing intermediate timescale searches, as well as including
an accurate description of the frequency evolution in a waveform model.

\begin{acknowledgments}
We thank Lavinia Heisenberg for useful discussions and comments on an earlier
version of this paper. We thank Chris Kavanagh and Peter Zimmerman for
advice and clarifications regarding the MST formalism. N.S. acknowledges
financial support by the German Academic Scholarship Foundation.  W.E.
acknowledges support from an NSERC Discovery grant.  Research at Perimeter
Institute is supported in part by the Government of Canada through the
Department of Innovation, Science and Economic Development Canada and by
the Province of Ontario through the Ministry of Economic Development, Job
Creation and Trade.  Calculations were performed on the Symmetry cluster at
Perimeter Institute.
\end{acknowledgments}

\appendix*
\section{Frequency fit functions}	 

In this appendix, we present our fit functions, and their corresponding accuracy,
for the real and imaginary parts of the frequency $\omega=\omega_R+i\omega_I$
of the superradiantly unstable Proca modes.
These were obtained by fitting to the numerical data obtained in Sec.~
\ref{procaresults}. We utilize the ansatz
\begin{align}
\begin{aligned}
\frac{\omega_R}{\mu}-1+\frac{\mu^2 M^2}{2m^2}= & \ (\mu M)^4 \hat{a}_{4,0,m} \\
& + \sum_{p\geq 5}\sum_{q\geq 0} (\mu M)^p \hat{a}_{p,q,m} (1-\chi^2)^{q/2}
\label{fitfuncR}
\end{aligned}
\end{align}
for the real part, and
\begin{align}
\begin{aligned}
\frac{\omega_I M(\mu M)^{-3-4m}}{C_m f_m(\omega_R-m\Omega_H)} -1=  \sum_{p\geq 1}\sum_{q\geq 0}(\mu M)^p\big[ \hat{b}_{p,q,m}\chi^{q+1} \\
 \hspace*{2cm}+\hat{c}_{p,q,m}(1-\chi^2)^{q/2}\big],
\label{fitfuncI}
\end{aligned}
\end{align}
for the imaginary part. Here $C_1=2$, $C_2=1/1728$, $C_3=4/(199290375)$, and
$f_m=\prod_{h=1}^{m}[h^2(1-\chi^2)+(\chi m-2r_+\omega_R)^2]$
\cite{Baumann:2019eav,Baryakhtar:2017ngi}, while $\chi=a/M$. We present fit
functions for the frequency of only the fundamental ($\hat{n}=0$) modes with $m=1$ 
through $m=3$, since these dominate the
dynamics for the largest part of the parameter space. In Eqs.~\eqref{fitfuncR} and
\eqref{fitfuncI}, we
factor out the leading-order contributions in the non-relativistic limit of
$\omega_R$ \cite{Baryakhtar:2017ngi}. As discussed below, we have tested these fits 
only against our numerical results in the relativistic regime. However, we factored out 
the known non-relativistic corrections, and therefore expect those fits to also 
do well in the rest of the parameter space. Notice, though, that there is a $\mu M$
range between the regime in which we tested the fits and the non-relativistic limit, 
where the fits may behave differently from the true result. This matching range
increases in size with increasing azimuthal mode number. The coefficients for the
real parts of the frequencies are given in \tablename{ \ref{coeff1}}, whereas the
coefficients for the imaginary parts are given in \tablename{ \ref{coeff23}}.  

\begin{table}[t]
\begin{ruledtabular}
\begin{tabular}{c|cccc}
$\hat{a}_{p,q,1}$ & $q=0$ & $q=1$ & $2$ & $3$ \\ \hline 
$p=5$ &  -0.870039 & 1.92247 & -6.93044 & -0.651989 \\
$6$ &  6.00734 & -10.4118 & 30.3691 & 18.2029 \\
$7$ & -4.23715 & 11.2671 & -33.6966 & -102.444 \\ \hline \hline
$\hat{a}_{p,q,2}$ & $q=0$ & $q=1$ & $2$ & $3$ \\ \hline 
$p=5$ &  -0.00276306 & 0.025062 & -0.164217 & 0.0216578 \\
$6$ &  0.0380878 & -0.0949728 & 0.458643 & -0.00599547 \\
$7$ & -0.0093238 & 0.0654495 & -0.333585 & -0.26639 \\ \hline \hline
$\hat{a}_{p,q,3}$ & $q=0$ & $q=1$ & $2$ & $3$ \\ \hline 
$p=0$ &  -0.000649075 & 0.00748510 & -0.0294351 & 0.00695179 \\
$6$ & 0.00362974 & -0.0131563 & 0.0434778 & -0.000717772 \\
$7$ & -0.000682509 & 0.00508197 & -0.0175571 & -0.0139853 \\
\end{tabular}
\end{ruledtabular}
\caption{The fit coefficients for the real part of the frequency, as defined in \eqref{fitfuncR} and \eqref{fitfuncI}. Additionally, $\hat{a}_{4,0,1}=-1.16304$, $\hat{a}_{4,0,2}=-0.0614401$ and $\hat{a}_{4,0,3}=-0.0116139$.}
\label{coeff1}
\end{table}

\begin{table*}[t]
\begin{ruledtabular}
\begin{tabular}{c|ccccccccccc}
$\hat{b}_{p,q,1}\times 10^{-5}$ & $p=1$ & $2$ & $3$ & $4$ & $5$ & $6$ & $7$ & $8$ & $9$ & $10$ \\ \hline
$q=0$ & -0.000172849 & 0.021852 & -0.957118 & 19.2205 & -189.031 & 794.509 & 319.628 & -14176. & 43620.8 & -42174.6 \\
$1$ & -0.000058719 & 0.00396914 & 0.0401875 & -3.77957 & 58.1935 & -318.606 & 40.1126 & 5746.01 & -19453.2 & 19763.2 \\ \hline
$\hat{b}_{p,q,2}\times 10^{-5}$ & $p=5$ & $6$ & $7$ & $8$ & $9$ & $10$ & $11$ & $12$ & $13$ & $14$ \\ \hline
$q=0$ & 14.4455 & -235.964 & 1808.47 & -8368.42 & 25191.5 & -50221. & 65445.7 & -53266.5 & 24374.1 & -4724.85 \\
$1$ & -10.1872 & 146.829 & -999.726 & 4224.45 & -12019.5 & 23314.2 & -30142. & 24612.9 & -11363.4 & 2227.57  \\ \hline
$\hat{b}_{p,q,3}$ & $p=13$ & $14$ & $15$ & $16$ & $17$ & $18$ & $19$ & $20$ & $21$ & $22$ \\ \hline
$q=1$ & 15037.966 & -7589.684 & -11073.655 & -3450.736 & 1195.4877 & 1572.5096 & 897.2985 & 820.2915 & 978.5094 & 472.0680 \\
$2$ & -17130.990 & 14309.731 & 5272.649 & 2467.256 & 1106.0694 & -760.0280 & -1706.9831 & -1308.4764 & -543.7076 & -550.2190 \\
\end{tabular}
\end{ruledtabular}
\ \\
\ \\
\begin{ruledtabular}
\begin{tabular}{c|ccccccccccc}
$\hat{c}_{p,q,1}\times 10^{-5}$ & $p=1$ & $2$ & $3$ & $4$ & $5$ & $6$ & $7$ & $8$ & $9$ & $10$ \\ \hline
$q=0$ & 0.000150374 & -0.0236152 & 0.891463 & -15.3662 & 131.805 & -485.664 & -321.996 & 8357.09 & -24098.5 & 22385.9 \\
$1$ & 0.0000910064 & 0.00369826 & -0.207125 & 3.99915 & -37.1508 & 186.221 & -514.875 & 753.039 & -504.345 & 103.496 \\ \hline
$\hat{c}_{p,q,2}\times 10^{-5}$ & $p=5$ & $6$ & $7$ & $8$ & $9$ & $10$ & $11$ & $12$ & $13$ & $14$ \\ \hline
$q=0$ & -3.5933 & 80.1423 & -756.209 & 3969.73 & -12810.3 & 26419.5 & -34878.1 & 28421.6 & -12939.2 & 2487.81 \\
$1$ & -3.71268 & 47.4052 & -258.085 & 782.508 & -1441.81 & 1643.4 & -1114.8 & 391.394 & -37.5528 & -8.74807  \\ \hline
$\hat{c}_{p,q,3}$ & $p=13$ & $14$ & $15$ & $16$ & $17$ & $18$ & $19$ & $20$ & $21$ & $22$ \\ \hline
$q=1$ & 0 & -1060.1836 & -3388.984 & 6247.957 & 1138.9217 & -2112.0303 & -2230.7653 & 4.619473 & 1812.5793 & -599.9215 \\
$2$ & 114566.94 & -376625.9 & 396163.0 & -55852.15 & -120416.34 & 9180.998 & 45541.30 & 4094.197 & -19781.675 & 4743.193 \\
\end{tabular}
\end{ruledtabular}
\caption{The fit coefficients for the imaginary parts of the frequencies as defined in \eqref{fitfuncR} and \eqref{fitfuncI}. Notice that we chose $\hat{b}_{p\leq 4,q,2}=\hat{c}_{p\leq 4,q,2}=0$, as well as $\hat{b}_{p\leq 12,q,3}=\hat{c}_{p\leq 12,q,3}=0$, to find a good match of the non-relativistic limit with our results.}
\label{coeff23}
\end{table*}

We have compared the fit function for $m=1$ with our numerical data for the
same frequencies in the regime $\mu M\in (0.05,(\mu M)_\text{sat.})$, where
$(\mu M)_\text{sat.}$ is the saturation point for a given spin. 
In this regime, and for $a/M\in (0.4,0.99)$, we find a maximum
relative residual, $x_\text{max}(\omega_{R,I}):=\max
(|\omega_{R,I}^\text{fit}-\omega_{R,I}^\text{numerical}|/\omega_{R,I}^\text{numerical})$, in $\omega_R$
of $x_\text{max}^\text{m=1}(\omega_{R})=7.4 \times 10^{-5}$, whereas the mean
relative residual, $x_\text{mean}(\omega_{R,I}):=\text{mean}
(|\omega_{R,I}^\text{fit}-\omega_{R,I}^\text{numerical}|/\omega_{R,I}^\text{numerical})$,
over this part of the parameter space is
$x_\text{mean}^\text{m=1}(\omega_{R})=2.4 \times 10^{-6}$. Similarly, for
$\omega_I$, the fit functions residual in the region $(\mu M, a)\in (0.05,(\mu
M)_\text{sat.}),(0.4,0.99))$, is given by
$x_\text{max}^\text{m=1}(\omega_{I})=9.7 \times 10^{-3}$ and
$x_\text{mean}^\text{m=1}(\omega_{I})=1.4 \times 10^{-3}$. Beyond this regime,
i.e., for smaller spins $a$, we have not tested this fit.

For the $m=2$ and $m=3$ frequencies, we compared the fit functions to our data
in the regimes $(\mu M,a)_{m=2}\in ((0.35,(\mu M)_\text{sat.}),(0.7,0.99))$ and
$(\mu M,a)_{m=3}\in ((0.8,(\mu M)_\text{sat.}),(0.8,0.99))$, respectively. In
their respective regimes, we find the accuracy of the fits to be
\begin{align*}
\begin{aligned}
x_\text{max}^{m=2}(\omega_R)= & \ 6.9 \times 10^{-5}, \qquad  & x_\text{mean}^{m=2}(\omega_R)=7.5 \times 10^{-6}, \\
x_\text{max}^{m=2}(\omega_I)= & \ 1.4 \times 10^{-2}, \qquad  & x_\text{mean}^{m=2}(\omega_I)=4.0 \times 10^{-3},
\end{aligned}
\end{align*}
and
\begin{align*}
\begin{aligned}
x_\text{max}^{m=3}(\omega_R)= & \ 3.9 \times 10^{-5}, \qquad  & x_\text{mean}^{m=3}(\omega_R)=3.1\times 10^{-6}, \\
x_\text{max}^{m=3}(\omega_I)= & \ 3.1 \times 10^{-1}, \qquad  & x_\text{mean}^{m=3}(\omega_I)=2.5 \times 10^{-2}.
\end{aligned}
\end{align*}

\bibliography{bib.bib,references}

\begin{thebibliography}{53}%
\makeatletter
\providecommand \@ifxundefined [1]{%
 \@ifx{#1\undefined}
}%
\providecommand \@ifnum [1]{%
 \ifnum #1\expandafter \@firstoftwo
 \else \expandafter \@secondoftwo
 \fi
}%
\providecommand \@ifx [1]{%
 \ifx #1\expandafter \@firstoftwo
 \else \expandafter \@secondoftwo
 \fi
}%
\providecommand \natexlab [1]{#1}%
\providecommand \enquote  [1]{``#1''}%
\providecommand \bibnamefont  [1]{#1}%
\providecommand \bibfnamefont [1]{#1}%
\providecommand \citenamefont [1]{#1}%
\providecommand \href@noop [0]{\@secondoftwo}%
\providecommand \href [0]{\begingroup \@sanitize@url \@href}%
\providecommand \@href[1]{\@@startlink{#1}\@@href}%
\providecommand \@@href[1]{\endgroup#1\@@endlink}%
\providecommand \@sanitize@url [0]{\catcode `\\12\catcode `\$12\catcode
  `\&12\catcode `\#12\catcode `\^12\catcode `\_12\catcode `\%12\relax}%
\providecommand \@@startlink[1]{}%
\providecommand \@@endlink[0]{}%
\providecommand \url  [0]{\begingroup\@sanitize@url \@url }%
\providecommand \@url [1]{\endgroup\@href {#1}{\urlprefix }}%
\providecommand \urlprefix  [0]{URL }%
\providecommand \Eprint [0]{\href }%
\providecommand \doibase [0]{http://dx.doi.org/}%
\providecommand \selectlanguage [0]{\@gobble}%
\providecommand \bibinfo  [0]{\@secondoftwo}%
\providecommand \bibfield  [0]{\@secondoftwo}%
\providecommand \translation [1]{[#1]}%
\providecommand \BibitemOpen [0]{}%
\providecommand \bibitemStop [0]{}%
\providecommand \bibitemNoStop [0]{.\EOS\space}%
\providecommand \EOS [0]{\spacefactor3000\relax}%
\providecommand \BibitemShut  [1]{\csname bibitem#1\endcsname}%
\let\auto@bib@innerbib\@empty
\bibitem [{\citenamefont {Arvanitaki}\ \emph {et~al.}(2010)\citenamefont
  {Arvanitaki}, \citenamefont {Dimopoulos}, \citenamefont {Dubovsky},
  \citenamefont {Kaloper},\ and\ \citenamefont
  {March-Russell}}]{Arvanitaki:2009fg}%
  \BibitemOpen
  \bibfield  {author} {\bibinfo {author} {\bibfnamefont {A.}~\bibnamefont
  {Arvanitaki}}, \bibinfo {author} {\bibfnamefont {S.}~\bibnamefont
  {Dimopoulos}}, \bibinfo {author} {\bibfnamefont {S.}~\bibnamefont
  {Dubovsky}}, \bibinfo {author} {\bibfnamefont {N.}~\bibnamefont {Kaloper}}, \
  and\ \bibinfo {author} {\bibfnamefont {J.}~\bibnamefont {March-Russell}},\
  }\href {\doibase 10.1103/PhysRevD.81.123530} {\bibfield  {journal} {\bibinfo
  {journal} {Phys. Rev.}\ }\textbf {\bibinfo {volume} {D81}},\ \bibinfo {pages}
  {123530} (\bibinfo {year} {2010})},\ \Eprint {http://arxiv.org/abs/0905.4720}
  {arXiv:0905.4720 [hep-th]} \BibitemShut {NoStop}%
\bibitem [{\citenamefont {Jaeckel}\ and\ \citenamefont
  {Ringwald}(2010)}]{Jaeckel:2010ni}%
  \BibitemOpen
  \bibfield  {author} {\bibinfo {author} {\bibfnamefont {J.}~\bibnamefont
  {Jaeckel}}\ and\ \bibinfo {author} {\bibfnamefont {A.}~\bibnamefont
  {Ringwald}},\ }\href {\doibase 10.1146/annurev.nucl.012809.104433} {\bibfield
   {journal} {\bibinfo  {journal} {Ann. Rev. Nucl. Part. Sci.}\ }\textbf
  {\bibinfo {volume} {60}},\ \bibinfo {pages} {405} (\bibinfo {year} {2010})},\
  \Eprint {http://arxiv.org/abs/1002.0329} {arXiv:1002.0329 [hep-ph]}
  \BibitemShut {NoStop}%
\bibitem [{\citenamefont {Arvanitaki}\ and\ \citenamefont
  {Dubovsky}(2011)}]{Arvanitaki:2010sy}%
  \BibitemOpen
  \bibfield  {author} {\bibinfo {author} {\bibfnamefont {A.}~\bibnamefont
  {Arvanitaki}}\ and\ \bibinfo {author} {\bibfnamefont {S.}~\bibnamefont
  {Dubovsky}},\ }\href {\doibase 10.1103/PhysRevD.83.044026} {\bibfield
  {journal} {\bibinfo  {journal} {Phys. Rev.}\ }\textbf {\bibinfo {volume}
  {D83}},\ \bibinfo {pages} {044026} (\bibinfo {year} {2011})},\ \Eprint
  {http://arxiv.org/abs/1004.3558} {arXiv:1004.3558 [hep-th]} \BibitemShut
  {NoStop}%
\bibitem [{\citenamefont {Agrawal}\ \emph {et~al.}(2018)\citenamefont
  {Agrawal}, \citenamefont {Kitajima}, \citenamefont {Reece}, \citenamefont
  {Sekiguchi},\ and\ \citenamefont {Takahashi}}]{Agrawal:2018vin}%
  \BibitemOpen
  \bibfield  {author} {\bibinfo {author} {\bibfnamefont {P.}~\bibnamefont
  {Agrawal}}, \bibinfo {author} {\bibfnamefont {N.}~\bibnamefont {Kitajima}},
  \bibinfo {author} {\bibfnamefont {M.}~\bibnamefont {Reece}}, \bibinfo
  {author} {\bibfnamefont {T.}~\bibnamefont {Sekiguchi}}, \ and\ \bibinfo
  {author} {\bibfnamefont {F.}~\bibnamefont {Takahashi}},\ }\href@noop {} {\
  (\bibinfo {year} {2018})},\ \Eprint {http://arxiv.org/abs/1810.07188}
  {arXiv:1810.07188 [hep-ph]} \BibitemShut {NoStop}%
\bibitem [{\citenamefont {Essig}\ \emph {et~al.}(2013)\citenamefont {Essig}
  \emph {et~al.}}]{Essig:2013lka}%
  \BibitemOpen
  \bibfield  {author} {\bibinfo {author} {\bibfnamefont {R.}~\bibnamefont
  {Essig}} \emph {et~al.},\ }in\ \href
  {http://www.slac.stanford.edu/econf/C1307292/docs/IntensityFrontier/NewLight-17.pdf}
  {\emph {\bibinfo {booktitle} {{Proceedings, 2013 Community Summer Study on
  the Future of U.S. Particle Physics: Snowmass on the Mississippi (CSS2013):
  Minneapolis, MN, USA, July 29-August 6, 2013}}}}\ (\bibinfo {year} {2013})\
  \Eprint {http://arxiv.org/abs/1311.0029} {arXiv:1311.0029 [hep-ph]}
  \BibitemShut {NoStop}%
\bibitem [{\citenamefont {De~Felice}\ \emph
  {et~al.}(2016{\natexlab{a}})\citenamefont {De~Felice}, \citenamefont
  {Heisenberg}, \citenamefont {Kase}, \citenamefont {Mukohyama}, \citenamefont
  {Tsujikawa},\ and\ \citenamefont {Zhang}}]{DeFelice:2016uil}%
  \BibitemOpen
  \bibfield  {author} {\bibinfo {author} {\bibfnamefont {A.}~\bibnamefont
  {De~Felice}}, \bibinfo {author} {\bibfnamefont {L.}~\bibnamefont
  {Heisenberg}}, \bibinfo {author} {\bibfnamefont {R.}~\bibnamefont {Kase}},
  \bibinfo {author} {\bibfnamefont {S.}~\bibnamefont {Mukohyama}}, \bibinfo
  {author} {\bibfnamefont {S.}~\bibnamefont {Tsujikawa}}, \ and\ \bibinfo
  {author} {\bibfnamefont {Y.-l.}\ \bibnamefont {Zhang}},\ }\href {\doibase
  10.1103/PhysRevD.94.044024} {\bibfield  {journal} {\bibinfo  {journal} {Phys.
  Rev.}\ }\textbf {\bibinfo {volume} {D94}},\ \bibinfo {pages} {044024}
  (\bibinfo {year} {2016}{\natexlab{a}})},\ \Eprint
  {http://arxiv.org/abs/1605.05066} {arXiv:1605.05066 [gr-qc]} \BibitemShut
  {NoStop}%
\bibitem [{\citenamefont {De~Felice}\ \emph
  {et~al.}(2016{\natexlab{b}})\citenamefont {De~Felice}, \citenamefont
  {Heisenberg}, \citenamefont {Kase}, \citenamefont {Mukohyama}, \citenamefont
  {Tsujikawa},\ and\ \citenamefont {Zhang}}]{DeFelice:2016yws}%
  \BibitemOpen
  \bibfield  {author} {\bibinfo {author} {\bibfnamefont {A.}~\bibnamefont
  {De~Felice}}, \bibinfo {author} {\bibfnamefont {L.}~\bibnamefont
  {Heisenberg}}, \bibinfo {author} {\bibfnamefont {R.}~\bibnamefont {Kase}},
  \bibinfo {author} {\bibfnamefont {S.}~\bibnamefont {Mukohyama}}, \bibinfo
  {author} {\bibfnamefont {S.}~\bibnamefont {Tsujikawa}}, \ and\ \bibinfo
  {author} {\bibfnamefont {Y.-l.}\ \bibnamefont {Zhang}},\ }\href {\doibase
  10.1088/1475-7516/2016/06/048} {\bibfield  {journal} {\bibinfo  {journal}
  {JCAP}\ }\textbf {\bibinfo {volume} {1606}},\ \bibinfo {pages} {048}
  (\bibinfo {year} {2016}{\natexlab{b}})},\ \Eprint
  {http://arxiv.org/abs/1603.05806} {arXiv:1603.05806 [gr-qc]} \BibitemShut
  {NoStop}%
\bibitem [{\citenamefont {Peccei}\ and\ \citenamefont
  {Quinn}(1977)}]{Peccei:1977hh}%
  \BibitemOpen
  \bibfield  {author} {\bibinfo {author} {\bibfnamefont {R.~D.}\ \bibnamefont
  {Peccei}}\ and\ \bibinfo {author} {\bibfnamefont {H.~R.}\ \bibnamefont
  {Quinn}},\ }\href {\doibase 10.1103/PhysRevLett.38.1440} {\bibfield
  {journal} {\bibinfo  {journal} {Phys. Rev. Lett.}\ }\textbf {\bibinfo
  {volume} {38}},\ \bibinfo {pages} {1440} (\bibinfo {year}
  {1977})}\BibitemShut {NoStop}%
\bibitem [{\citenamefont {Weinberg}(1978)}]{Weinberg:1977ma}%
  \BibitemOpen
  \bibfield  {author} {\bibinfo {author} {\bibfnamefont {S.}~\bibnamefont
  {Weinberg}},\ }\href {\doibase 10.1103/PhysRevLett.40.223} {\bibfield
  {journal} {\bibinfo  {journal} {Phys. Rev. Lett.}\ }\textbf {\bibinfo
  {volume} {40}},\ \bibinfo {pages} {223} (\bibinfo {year} {1978})}\BibitemShut
  {NoStop}%
\bibitem [{\citenamefont {Abbott}\ \emph {et~al.}(2016)\citenamefont {Abbott}
  \emph {et~al.}}]{Abbott:2016nmj}%
  \BibitemOpen
  \bibfield  {author} {\bibinfo {author} {\bibfnamefont {B.~P.}\ \bibnamefont
  {Abbott}} \emph {et~al.} (\bibinfo {collaboration} {LIGO Scientific,
  Virgo}),\ }\href {\doibase 10.1103/PhysRevLett.116.241103} {\bibfield
  {journal} {\bibinfo  {journal} {Phys. Rev. Lett.}\ }\textbf {\bibinfo
  {volume} {116}},\ \bibinfo {pages} {241103} (\bibinfo {year} {2016})},\
  \Eprint {http://arxiv.org/abs/1606.04855} {arXiv:1606.04855 [gr-qc]}
  \BibitemShut {NoStop}%
\bibitem [{\citenamefont {Abbott}\ \emph {et~al.}(2019)\citenamefont {Abbott}
  \emph {et~al.}}]{LIGOScientific:2018mvr}%
  \BibitemOpen
  \bibfield  {author} {\bibinfo {author} {\bibfnamefont {B.~P.}\ \bibnamefont
  {Abbott}} \emph {et~al.} (\bibinfo {collaboration} {LIGO Scientific,
  Virgo}),\ }\href {\doibase 10.1103/PhysRevX.9.031040} {\bibfield  {journal}
  {\bibinfo  {journal} {Phys. Rev.}\ }\textbf {\bibinfo {volume} {X9}},\
  \bibinfo {pages} {031040} (\bibinfo {year} {2019})},\ \Eprint
  {http://arxiv.org/abs/1811.12907} {arXiv:1811.12907 [astro-ph.HE]}
  \BibitemShut {NoStop}%
\bibitem [{\citenamefont {Bardeen}\ \emph {et~al.}(1973)\citenamefont
  {Bardeen}, \citenamefont {Carter},\ and\ \citenamefont
  {Hawking}}]{Bardeen:1973gs}%
  \BibitemOpen
  \bibfield  {author} {\bibinfo {author} {\bibfnamefont {J.~M.}\ \bibnamefont
  {Bardeen}}, \bibinfo {author} {\bibfnamefont {B.}~\bibnamefont {Carter}}, \
  and\ \bibinfo {author} {\bibfnamefont {S.~W.}\ \bibnamefont {Hawking}},\
  }\href {\doibase 10.1007/BF01645742} {\bibfield  {journal} {\bibinfo
  {journal} {Commun. Math. Phys.}\ }\textbf {\bibinfo {volume} {31}},\ \bibinfo
  {pages} {161} (\bibinfo {year} {1973})}\BibitemShut {NoStop}%
\bibitem [{\citenamefont {Press}\ and\ \citenamefont
  {Teukolsky}(1972)}]{Press:1972zz}%
  \BibitemOpen
  \bibfield  {author} {\bibinfo {author} {\bibfnamefont {W.~H.}\ \bibnamefont
  {Press}}\ and\ \bibinfo {author} {\bibfnamefont {S.~A.}\ \bibnamefont
  {Teukolsky}},\ }\href {\doibase 10.1038/238211a0} {\bibfield  {journal}
  {\bibinfo  {journal} {Nature}\ }\textbf {\bibinfo {volume} {238}},\ \bibinfo
  {pages} {211} (\bibinfo {year} {1972})}\BibitemShut {NoStop}%
\bibitem [{\citenamefont {East}(2018)}]{East:2018glu}%
  \BibitemOpen
  \bibfield  {author} {\bibinfo {author} {\bibfnamefont {W.~E.}\ \bibnamefont
  {East}},\ }\href {\doibase 10.1103/PhysRevLett.121.131104} {\bibfield
  {journal} {\bibinfo  {journal} {Phys. Rev. Lett.}\ }\textbf {\bibinfo
  {volume} {121}},\ \bibinfo {pages} {131104} (\bibinfo {year} {2018})},\
  \Eprint {http://arxiv.org/abs/1807.00043} {arXiv:1807.00043 [gr-qc]}
  \BibitemShut {NoStop}%
\bibitem [{\citenamefont {East}\ and\ \citenamefont
  {Pretorius}(2017)}]{East:2017ovw}%
  \BibitemOpen
  \bibfield  {author} {\bibinfo {author} {\bibfnamefont {W.~E.}\ \bibnamefont
  {East}}\ and\ \bibinfo {author} {\bibfnamefont {F.}~\bibnamefont
  {Pretorius}},\ }\href {\doibase 10.1103/PhysRevLett.119.041101} {\bibfield
  {journal} {\bibinfo  {journal} {Phys. Rev. Lett.}\ }\textbf {\bibinfo
  {volume} {119}},\ \bibinfo {pages} {041101} (\bibinfo {year} {2017})},\
  \Eprint {http://arxiv.org/abs/1704.04791} {arXiv:1704.04791 [gr-qc]}
  \BibitemShut {NoStop}%
\bibitem [{\citenamefont {Davoudiasl}\ and\ \citenamefont
  {Denton}(2019)}]{Davoudiasl:2019nlo}%
  \BibitemOpen
  \bibfield  {author} {\bibinfo {author} {\bibfnamefont {H.}~\bibnamefont
  {Davoudiasl}}\ and\ \bibinfo {author} {\bibfnamefont {P.~B.}\ \bibnamefont
  {Denton}},\ }\href@noop {} {\  (\bibinfo {year} {2019})},\ \Eprint
  {http://arxiv.org/abs/1904.09242} {arXiv:1904.09242 [astro-ph.CO]}
  \BibitemShut {NoStop}%
\bibitem [{\citenamefont {Cardoso}\ \emph {et~al.}(2018)\citenamefont
  {Cardoso}, \citenamefont {Dias}, \citenamefont {Hartnett}, \citenamefont
  {Middleton}, \citenamefont {Pani},\ and\ \citenamefont
  {Santos}}]{Cardoso:2018tly}%
  \BibitemOpen
  \bibfield  {author} {\bibinfo {author} {\bibfnamefont {V.}~\bibnamefont
  {Cardoso}}, \bibinfo {author} {\bibfnamefont {O.~J.~C.}\ \bibnamefont
  {Dias}}, \bibinfo {author} {\bibfnamefont {G.~S.}\ \bibnamefont {Hartnett}},
  \bibinfo {author} {\bibfnamefont {M.}~\bibnamefont {Middleton}}, \bibinfo
  {author} {\bibfnamefont {P.}~\bibnamefont {Pani}}, \ and\ \bibinfo {author}
  {\bibfnamefont {J.~E.}\ \bibnamefont {Santos}},\ }\href {\doibase
  10.1088/1475-7516/2018/03/043} {\bibfield  {journal} {\bibinfo  {journal}
  {JCAP}\ }\textbf {\bibinfo {volume} {1803}},\ \bibinfo {pages} {043}
  (\bibinfo {year} {2018})},\ \Eprint {http://arxiv.org/abs/1801.01420}
  {arXiv:1801.01420 [gr-qc]} \BibitemShut {NoStop}%
\bibitem [{\citenamefont {Baumann}\ \emph
  {et~al.}(2019{\natexlab{a}})\citenamefont {Baumann}, \citenamefont {Chia},\
  and\ \citenamefont {Porto}}]{Baumann:2018vus}%
  \BibitemOpen
  \bibfield  {author} {\bibinfo {author} {\bibfnamefont {D.}~\bibnamefont
  {Baumann}}, \bibinfo {author} {\bibfnamefont {H.~S.}\ \bibnamefont {Chia}}, \
  and\ \bibinfo {author} {\bibfnamefont {R.~A.}\ \bibnamefont {Porto}},\ }\href
  {\doibase 10.1103/PhysRevD.99.044001} {\bibfield  {journal} {\bibinfo
  {journal} {Phys. Rev.}\ }\textbf {\bibinfo {volume} {D99}},\ \bibinfo {pages}
  {044001} (\bibinfo {year} {2019}{\natexlab{a}})},\ \Eprint
  {http://arxiv.org/abs/1804.03208} {arXiv:1804.03208 [gr-qc]} \BibitemShut
  {NoStop}%
\bibitem [{\citenamefont {Berti}\ \emph {et~al.}(2019)\citenamefont {Berti},
  \citenamefont {Brito}, \citenamefont {Macedo}, \citenamefont {Raposo},\ and\
  \citenamefont {Rosa}}]{Berti:2019wnn}%
  \BibitemOpen
  \bibfield  {author} {\bibinfo {author} {\bibfnamefont {E.}~\bibnamefont
  {Berti}}, \bibinfo {author} {\bibfnamefont {R.}~\bibnamefont {Brito}},
  \bibinfo {author} {\bibfnamefont {C.~F.~B.}\ \bibnamefont {Macedo}}, \bibinfo
  {author} {\bibfnamefont {G.}~\bibnamefont {Raposo}}, \ and\ \bibinfo {author}
  {\bibfnamefont {J.~L.}\ \bibnamefont {Rosa}},\ }\href {\doibase
  10.1103/PhysRevD.99.104039} {\bibfield  {journal} {\bibinfo  {journal} {Phys.
  Rev.}\ }\textbf {\bibinfo {volume} {D99}},\ \bibinfo {pages} {104039}
  (\bibinfo {year} {2019})},\ \Eprint {http://arxiv.org/abs/1904.03131}
  {arXiv:1904.03131 [gr-qc]} \BibitemShut {NoStop}%
\bibitem [{\citenamefont {Zhang}\ and\ \citenamefont
  {Yang}(2019{\natexlab{a}})}]{Zhang:2018kib}%
  \BibitemOpen
  \bibfield  {author} {\bibinfo {author} {\bibfnamefont {J.}~\bibnamefont
  {Zhang}}\ and\ \bibinfo {author} {\bibfnamefont {H.}~\bibnamefont {Yang}},\
  }\href {\doibase 10.1103/PhysRevD.99.064018} {\bibfield  {journal} {\bibinfo
  {journal} {Phys. Rev.}\ }\textbf {\bibinfo {volume} {D99}},\ \bibinfo {pages}
  {064018} (\bibinfo {year} {2019}{\natexlab{a}})},\ \Eprint
  {http://arxiv.org/abs/1808.02905} {arXiv:1808.02905 [gr-qc]} \BibitemShut
  {NoStop}%
\bibitem [{\citenamefont {Zhang}\ and\ \citenamefont
  {Yang}(2019{\natexlab{b}})}]{Zhang:2019eid}%
  \BibitemOpen
  \bibfield  {author} {\bibinfo {author} {\bibfnamefont {J.}~\bibnamefont
  {Zhang}}\ and\ \bibinfo {author} {\bibfnamefont {H.}~\bibnamefont {Yang}},\
  }\href@noop {} {\  (\bibinfo {year} {2019}{\natexlab{b}})},\ \Eprint
  {http://arxiv.org/abs/1907.13582} {arXiv:1907.13582 [gr-qc]} \BibitemShut
  {NoStop}%
\bibitem [{\citenamefont {Herdeiro}\ \emph {et~al.}(2016)\citenamefont
  {Herdeiro}, \citenamefont {Radu},\ and\ \citenamefont
  {Rúnarsson}}]{Herdeiro:2016tmi}%
  \BibitemOpen
  \bibfield  {author} {\bibinfo {author} {\bibfnamefont {C.}~\bibnamefont
  {Herdeiro}}, \bibinfo {author} {\bibfnamefont {E.}~\bibnamefont {Radu}}, \
  and\ \bibinfo {author} {\bibfnamefont {H.}~\bibnamefont {Rúnarsson}},\
  }\href {\doibase 10.1088/0264-9381/33/15/154001} {\bibfield  {journal}
  {\bibinfo  {journal} {Class. Quant. Grav.}\ }\textbf {\bibinfo {volume}
  {33}},\ \bibinfo {pages} {154001} (\bibinfo {year} {2016})},\ \Eprint
  {http://arxiv.org/abs/1603.02687} {arXiv:1603.02687 [gr-qc]} \BibitemShut
  {NoStop}%
\bibitem [{\citenamefont {Herdeiro}\ and\ \citenamefont
  {Radu}(2017)}]{Herdeiro:2017phl}%
  \BibitemOpen
  \bibfield  {author} {\bibinfo {author} {\bibfnamefont {C.~A.~R.}\
  \bibnamefont {Herdeiro}}\ and\ \bibinfo {author} {\bibfnamefont
  {E.}~\bibnamefont {Radu}},\ }\href {\doibase 10.1103/PhysRevLett.119.261101}
  {\bibfield  {journal} {\bibinfo  {journal} {Phys. Rev. Lett.}\ }\textbf
  {\bibinfo {volume} {119}},\ \bibinfo {pages} {261101} (\bibinfo {year}
  {2017})},\ \Eprint {http://arxiv.org/abs/1706.06597} {arXiv:1706.06597
  [gr-qc]} \BibitemShut {NoStop}%
\bibitem [{\citenamefont {Ganchev}\ and\ \citenamefont
  {Santos}(2018)}]{Ganchev:2017uuo}%
  \BibitemOpen
  \bibfield  {author} {\bibinfo {author} {\bibfnamefont {B.}~\bibnamefont
  {Ganchev}}\ and\ \bibinfo {author} {\bibfnamefont {J.~E.}\ \bibnamefont
  {Santos}},\ }\href {\doibase 10.1103/PhysRevLett.120.171101} {\bibfield
  {journal} {\bibinfo  {journal} {Phys. Rev. Lett.}\ }\textbf {\bibinfo
  {volume} {120}},\ \bibinfo {pages} {171101} (\bibinfo {year} {2018})},\
  \Eprint {http://arxiv.org/abs/1711.08464} {arXiv:1711.08464 [gr-qc]}
  \BibitemShut {NoStop}%
\bibitem [{\citenamefont {Arvanitaki}\ \emph {et~al.}(2015)\citenamefont
  {Arvanitaki}, \citenamefont {Baryakhtar},\ and\ \citenamefont
  {Huang}}]{Arvanitaki:2014wva}%
  \BibitemOpen
  \bibfield  {author} {\bibinfo {author} {\bibfnamefont {A.}~\bibnamefont
  {Arvanitaki}}, \bibinfo {author} {\bibfnamefont {M.}~\bibnamefont
  {Baryakhtar}}, \ and\ \bibinfo {author} {\bibfnamefont {X.}~\bibnamefont
  {Huang}},\ }\href {\doibase 10.1103/PhysRevD.91.084011} {\bibfield  {journal}
  {\bibinfo  {journal} {Phys. Rev.}\ }\textbf {\bibinfo {volume} {D91}},\
  \bibinfo {pages} {084011} (\bibinfo {year} {2015})},\ \Eprint
  {http://arxiv.org/abs/1411.2263} {arXiv:1411.2263 [hep-ph]} \BibitemShut
  {NoStop}%
\bibitem [{\citenamefont {Arvanitaki}\ \emph {et~al.}(2017)\citenamefont
  {Arvanitaki}, \citenamefont {Baryakhtar}, \citenamefont {Dimopoulos},
  \citenamefont {Dubovsky},\ and\ \citenamefont
  {Lasenby}}]{Arvanitaki:2016qwi}%
  \BibitemOpen
  \bibfield  {author} {\bibinfo {author} {\bibfnamefont {A.}~\bibnamefont
  {Arvanitaki}}, \bibinfo {author} {\bibfnamefont {M.}~\bibnamefont
  {Baryakhtar}}, \bibinfo {author} {\bibfnamefont {S.}~\bibnamefont
  {Dimopoulos}}, \bibinfo {author} {\bibfnamefont {S.}~\bibnamefont
  {Dubovsky}}, \ and\ \bibinfo {author} {\bibfnamefont {R.}~\bibnamefont
  {Lasenby}},\ }\href {\doibase 10.1103/PhysRevD.95.043001} {\bibfield
  {journal} {\bibinfo  {journal} {Phys. Rev.}\ }\textbf {\bibinfo {volume}
  {D95}},\ \bibinfo {pages} {043001} (\bibinfo {year} {2017})},\ \Eprint
  {http://arxiv.org/abs/1604.03958} {arXiv:1604.03958 [hep-ph]} \BibitemShut
  {NoStop}%
\bibitem [{\citenamefont {Tsukada}\ \emph {et~al.}(2019)\citenamefont
  {Tsukada}, \citenamefont {Callister}, \citenamefont {Matas},\ and\
  \citenamefont {Meyers}}]{Tsukada:2018mbp}%
  \BibitemOpen
  \bibfield  {author} {\bibinfo {author} {\bibfnamefont {L.}~\bibnamefont
  {Tsukada}}, \bibinfo {author} {\bibfnamefont {T.}~\bibnamefont {Callister}},
  \bibinfo {author} {\bibfnamefont {A.}~\bibnamefont {Matas}}, \ and\ \bibinfo
  {author} {\bibfnamefont {P.}~\bibnamefont {Meyers}},\ }\href {\doibase
  10.1103/PhysRevD.99.103015} {\bibfield  {journal} {\bibinfo  {journal} {Phys.
  Rev.}\ }\textbf {\bibinfo {volume} {D99}},\ \bibinfo {pages} {103015}
  (\bibinfo {year} {2019})},\ \Eprint {http://arxiv.org/abs/1812.09622}
  {arXiv:1812.09622 [astro-ph.HE]} \BibitemShut {NoStop}%
\bibitem [{\citenamefont {Brito}\ \emph
  {et~al.}(2017{\natexlab{a}})\citenamefont {Brito}, \citenamefont {Ghosh},
  \citenamefont {Barausse}, \citenamefont {Berti}, \citenamefont {Cardoso},
  \citenamefont {Dvorkin}, \citenamefont {Klein},\ and\ \citenamefont
  {Pani}}]{Brito:2017wnc}%
  \BibitemOpen
  \bibfield  {author} {\bibinfo {author} {\bibfnamefont {R.}~\bibnamefont
  {Brito}}, \bibinfo {author} {\bibfnamefont {S.}~\bibnamefont {Ghosh}},
  \bibinfo {author} {\bibfnamefont {E.}~\bibnamefont {Barausse}}, \bibinfo
  {author} {\bibfnamefont {E.}~\bibnamefont {Berti}}, \bibinfo {author}
  {\bibfnamefont {V.}~\bibnamefont {Cardoso}}, \bibinfo {author} {\bibfnamefont
  {I.}~\bibnamefont {Dvorkin}}, \bibinfo {author} {\bibfnamefont
  {A.}~\bibnamefont {Klein}}, \ and\ \bibinfo {author} {\bibfnamefont
  {P.}~\bibnamefont {Pani}},\ }\href {\doibase 10.1103/PhysRevLett.119.131101}
  {\bibfield  {journal} {\bibinfo  {journal} {Phys. Rev. Lett.}\ }\textbf
  {\bibinfo {volume} {119}},\ \bibinfo {pages} {131101} (\bibinfo {year}
  {2017}{\natexlab{a}})},\ \Eprint {http://arxiv.org/abs/1706.05097}
  {arXiv:1706.05097 [gr-qc]} \BibitemShut {NoStop}%
\bibitem [{\citenamefont {Brito}\ \emph
  {et~al.}(2017{\natexlab{b}})\citenamefont {Brito}, \citenamefont {Ghosh},
  \citenamefont {Barausse}, \citenamefont {Berti}, \citenamefont {Cardoso},
  \citenamefont {Dvorkin}, \citenamefont {Klein},\ and\ \citenamefont
  {Pani}}]{Brito:2017zvb}%
  \BibitemOpen
  \bibfield  {author} {\bibinfo {author} {\bibfnamefont {R.}~\bibnamefont
  {Brito}}, \bibinfo {author} {\bibfnamefont {S.}~\bibnamefont {Ghosh}},
  \bibinfo {author} {\bibfnamefont {E.}~\bibnamefont {Barausse}}, \bibinfo
  {author} {\bibfnamefont {E.}~\bibnamefont {Berti}}, \bibinfo {author}
  {\bibfnamefont {V.}~\bibnamefont {Cardoso}}, \bibinfo {author} {\bibfnamefont
  {I.}~\bibnamefont {Dvorkin}}, \bibinfo {author} {\bibfnamefont
  {A.}~\bibnamefont {Klein}}, \ and\ \bibinfo {author} {\bibfnamefont
  {P.}~\bibnamefont {Pani}},\ }\href {\doibase 10.1103/PhysRevD.96.064050}
  {\bibfield  {journal} {\bibinfo  {journal} {Phys. Rev.}\ }\textbf {\bibinfo
  {volume} {D96}},\ \bibinfo {pages} {064050} (\bibinfo {year}
  {2017}{\natexlab{b}})},\ \Eprint {http://arxiv.org/abs/1706.06311}
  {arXiv:1706.06311 [gr-qc]} \BibitemShut {NoStop}%
\bibitem [{\citenamefont {Baryakhtar}\ \emph {et~al.}(2017)\citenamefont
  {Baryakhtar}, \citenamefont {Lasenby},\ and\ \citenamefont
  {Teo}}]{Baryakhtar:2017ngi}%
  \BibitemOpen
  \bibfield  {author} {\bibinfo {author} {\bibfnamefont {M.}~\bibnamefont
  {Baryakhtar}}, \bibinfo {author} {\bibfnamefont {R.}~\bibnamefont {Lasenby}},
  \ and\ \bibinfo {author} {\bibfnamefont {M.}~\bibnamefont {Teo}},\ }\href
  {\doibase 10.1103/PhysRevD.96.035019} {\bibfield  {journal} {\bibinfo
  {journal} {Phys. Rev.}\ }\textbf {\bibinfo {volume} {D96}},\ \bibinfo {pages}
  {035019} (\bibinfo {year} {2017})},\ \Eprint
  {http://arxiv.org/abs/1704.05081} {arXiv:1704.05081 [hep-ph]} \BibitemShut
  {NoStop}%
\bibitem [{\citenamefont {Baumann}\ \emph
  {et~al.}(2019{\natexlab{b}})\citenamefont {Baumann}, \citenamefont {Chia},
  \citenamefont {Stout},\ and\ \citenamefont {ter Haar}}]{Baumann:2019eav}%
  \BibitemOpen
  \bibfield  {author} {\bibinfo {author} {\bibfnamefont {D.}~\bibnamefont
  {Baumann}}, \bibinfo {author} {\bibfnamefont {H.~S.}\ \bibnamefont {Chia}},
  \bibinfo {author} {\bibfnamefont {J.}~\bibnamefont {Stout}}, \ and\ \bibinfo
  {author} {\bibfnamefont {L.}~\bibnamefont {ter Haar}},\ }\href@noop {} {\
  (\bibinfo {year} {2019}{\natexlab{b}})},\ \Eprint
  {http://arxiv.org/abs/1908.10370} {arXiv:1908.10370 [gr-qc]} \BibitemShut
  {NoStop}%
\bibitem [{\citenamefont {Pani}\ \emph {et~al.}(2012)\citenamefont {Pani},
  \citenamefont {Cardoso}, \citenamefont {Gualtieri}, \citenamefont {Berti},\
  and\ \citenamefont {Ishibashi}}]{Pani:2012bp}%
  \BibitemOpen
  \bibfield  {author} {\bibinfo {author} {\bibfnamefont {P.}~\bibnamefont
  {Pani}}, \bibinfo {author} {\bibfnamefont {V.}~\bibnamefont {Cardoso}},
  \bibinfo {author} {\bibfnamefont {L.}~\bibnamefont {Gualtieri}}, \bibinfo
  {author} {\bibfnamefont {E.}~\bibnamefont {Berti}}, \ and\ \bibinfo {author}
  {\bibfnamefont {A.}~\bibnamefont {Ishibashi}},\ }\href {\doibase
  10.1103/PhysRevD.86.104017} {\bibfield  {journal} {\bibinfo  {journal} {Phys.
  Rev.}\ }\textbf {\bibinfo {volume} {D86}},\ \bibinfo {pages} {104017}
  (\bibinfo {year} {2012})},\ \Eprint {http://arxiv.org/abs/1209.0773}
  {arXiv:1209.0773 [gr-qc]} \BibitemShut {NoStop}%
\bibitem [{\citenamefont {Witek}\ \emph {et~al.}(2013)\citenamefont {Witek},
  \citenamefont {Cardoso}, \citenamefont {Ishibashi},\ and\ \citenamefont
  {Sperhake}}]{Witek:2012tr}%
  \BibitemOpen
  \bibfield  {author} {\bibinfo {author} {\bibfnamefont {H.}~\bibnamefont
  {Witek}}, \bibinfo {author} {\bibfnamefont {V.}~\bibnamefont {Cardoso}},
  \bibinfo {author} {\bibfnamefont {A.}~\bibnamefont {Ishibashi}}, \ and\
  \bibinfo {author} {\bibfnamefont {U.}~\bibnamefont {Sperhake}},\ }\href
  {\doibase 10.1103/PhysRevD.87.043513} {\bibfield  {journal} {\bibinfo
  {journal} {Phys. Rev.}\ }\textbf {\bibinfo {volume} {D87}},\ \bibinfo {pages}
  {043513} (\bibinfo {year} {2013})},\ \Eprint {http://arxiv.org/abs/1212.0551}
  {arXiv:1212.0551 [gr-qc]} \BibitemShut {NoStop}%
\bibitem [{\citenamefont {East}(2017)}]{East:2017mrj}%
  \BibitemOpen
  \bibfield  {author} {\bibinfo {author} {\bibfnamefont {W.~E.}\ \bibnamefont
  {East}},\ }\href {\doibase 10.1103/PhysRevD.96.024004} {\bibfield  {journal}
  {\bibinfo  {journal} {Phys. Rev.}\ }\textbf {\bibinfo {volume} {D96}},\
  \bibinfo {pages} {024004} (\bibinfo {year} {2017})},\ \Eprint
  {http://arxiv.org/abs/1705.01544} {arXiv:1705.01544 [gr-qc]} \BibitemShut
  {NoStop}%
\bibitem [{\citenamefont {Frolov}\ \emph {et~al.}(2018)\citenamefont {Frolov},
  \citenamefont {Krtouš}, \citenamefont {Kubizňák},\ and\ \citenamefont
  {Santos}}]{Frolov:2018ezx}%
  \BibitemOpen
  \bibfield  {author} {\bibinfo {author} {\bibfnamefont {V.~P.}\ \bibnamefont
  {Frolov}}, \bibinfo {author} {\bibfnamefont {P.}~\bibnamefont {Krtouš}},
  \bibinfo {author} {\bibfnamefont {D.}~\bibnamefont {Kubizňák}}, \ and\
  \bibinfo {author} {\bibfnamefont {J.~E.}\ \bibnamefont {Santos}},\ }\href
  {\doibase 10.1103/PhysRevLett.120.231103} {\bibfield  {journal} {\bibinfo
  {journal} {Phys. Rev. Lett.}\ }\textbf {\bibinfo {volume} {120}},\ \bibinfo
  {pages} {231103} (\bibinfo {year} {2018})},\ \Eprint
  {http://arxiv.org/abs/1804.00030} {arXiv:1804.00030 [hep-th]} \BibitemShut
  {NoStop}%
\bibitem [{\citenamefont {Lunin}(2017)}]{Lunin:2017drx}%
  \BibitemOpen
  \bibfield  {author} {\bibinfo {author} {\bibfnamefont {O.}~\bibnamefont
  {Lunin}},\ }\href {\doibase 10.1007/JHEP12(2017)138} {\bibfield  {journal}
  {\bibinfo  {journal} {JHEP}\ }\textbf {\bibinfo {volume} {12}},\ \bibinfo
  {pages} {138} (\bibinfo {year} {2017})},\ \Eprint
  {http://arxiv.org/abs/1708.06766} {arXiv:1708.06766 [hep-th]} \BibitemShut
  {NoStop}%
\bibitem [{\citenamefont {Krtouš}\ \emph {et~al.}(2018)\citenamefont
  {Krtouš}, \citenamefont {Frolov},\ and\ \citenamefont
  {Kubizňák}}]{Krtous:2018bvk}%
  \BibitemOpen
  \bibfield  {author} {\bibinfo {author} {\bibfnamefont {P.}~\bibnamefont
  {Krtouš}}, \bibinfo {author} {\bibfnamefont {V.~P.}\ \bibnamefont {Frolov}},
  \ and\ \bibinfo {author} {\bibfnamefont {D.}~\bibnamefont {Kubizňák}},\
  }\href {\doibase 10.1016/j.nuclphysb.2018.06.019} {\bibfield  {journal}
  {\bibinfo  {journal} {Nucl. Phys.}\ }\textbf {\bibinfo {volume} {B934}},\
  \bibinfo {pages} {7} (\bibinfo {year} {2018})},\ \Eprint
  {http://arxiv.org/abs/1803.02485} {arXiv:1803.02485 [hep-th]} \BibitemShut
  {NoStop}%
\bibitem [{\citenamefont {Teukolsky}(1973)}]{Teukolsky:1973ha}%
  \BibitemOpen
  \bibfield  {author} {\bibinfo {author} {\bibfnamefont {S.~A.}\ \bibnamefont
  {Teukolsky}},\ }\href {\doibase 10.1086/152444} {\bibfield  {journal}
  {\bibinfo  {journal} {Astrophys. J.}\ }\textbf {\bibinfo {volume} {185}},\
  \bibinfo {pages} {635} (\bibinfo {year} {1973})}\BibitemShut {NoStop}%
\bibitem [{\citenamefont {Press}\ and\ \citenamefont
  {Teukolsky}(1973)}]{Press:1973zz}%
  \BibitemOpen
  \bibfield  {author} {\bibinfo {author} {\bibfnamefont {W.~H.}\ \bibnamefont
  {Press}}\ and\ \bibinfo {author} {\bibfnamefont {S.~A.}\ \bibnamefont
  {Teukolsky}},\ }\href {\doibase 10.1086/152445} {\bibfield  {journal}
  {\bibinfo  {journal} {Astrophys. J.}\ }\textbf {\bibinfo {volume} {185}},\
  \bibinfo {pages} {649} (\bibinfo {year} {1973})}\BibitemShut {NoStop}%
\bibitem [{\citenamefont {Ficarra}\ \emph {et~al.}(2019)\citenamefont
  {Ficarra}, \citenamefont {Pani},\ and\ \citenamefont
  {Witek}}]{Ficarra:2018rfu}%
  \BibitemOpen
  \bibfield  {author} {\bibinfo {author} {\bibfnamefont {G.}~\bibnamefont
  {Ficarra}}, \bibinfo {author} {\bibfnamefont {P.}~\bibnamefont {Pani}}, \
  and\ \bibinfo {author} {\bibfnamefont {H.}~\bibnamefont {Witek}},\ }\href
  {\doibase 10.1103/PhysRevD.99.104019} {\bibfield  {journal} {\bibinfo
  {journal} {Phys. Rev.}\ }\textbf {\bibinfo {volume} {D99}},\ \bibinfo {pages}
  {104019} (\bibinfo {year} {2019})},\ \Eprint
  {http://arxiv.org/abs/1812.02758} {arXiv:1812.02758 [gr-qc]} \BibitemShut
  {NoStop}%
\bibitem [{\citenamefont {Dolan}(2018)}]{Dolan:2018dqv}%
  \BibitemOpen
  \bibfield  {author} {\bibinfo {author} {\bibfnamefont {S.~R.}\ \bibnamefont
  {Dolan}},\ }\href {\doibase 10.1103/PhysRevD.98.104006} {\bibfield  {journal}
  {\bibinfo  {journal} {Phys. Rev.}\ }\textbf {\bibinfo {volume} {D98}},\
  \bibinfo {pages} {104006} (\bibinfo {year} {2018})},\ \Eprint
  {http://arxiv.org/abs/1806.01604} {arXiv:1806.01604 [gr-qc]} \BibitemShut
  {NoStop}%
\bibitem [{\citenamefont {Chesler}\ and\ \citenamefont
  {Lowe}(2019)}]{Chesler:2018txn}%
  \BibitemOpen
  \bibfield  {author} {\bibinfo {author} {\bibfnamefont {P.~M.}\ \bibnamefont
  {Chesler}}\ and\ \bibinfo {author} {\bibfnamefont {D.~A.}\ \bibnamefont
  {Lowe}},\ }\href {\doibase 10.1103/PhysRevLett.122.181101} {\bibfield
  {journal} {\bibinfo  {journal} {Phys. Rev. Lett.}\ }\textbf {\bibinfo
  {volume} {122}},\ \bibinfo {pages} {181101} (\bibinfo {year} {2019})},\
  \Eprint {http://arxiv.org/abs/1801.09711} {arXiv:1801.09711 [gr-qc]}
  \BibitemShut {NoStop}%
\bibitem [{\citenamefont {Sasaki}\ and\ \citenamefont
  {Tagoshi}(2003)}]{Sasaki:2003xr}%
  \BibitemOpen
  \bibfield  {author} {\bibinfo {author} {\bibfnamefont {M.}~\bibnamefont
  {Sasaki}}\ and\ \bibinfo {author} {\bibfnamefont {H.}~\bibnamefont
  {Tagoshi}},\ }\href {\doibase 10.12942/lrr-2003-6} {\bibfield  {journal}
  {\bibinfo  {journal} {Living Rev. Rel.}\ }\textbf {\bibinfo {volume} {6}},\
  \bibinfo {pages} {6} (\bibinfo {year} {2003})},\ \Eprint
  {http://arxiv.org/abs/gr-qc/0306120} {arXiv:gr-qc/0306120 [gr-qc]}
  \BibitemShut {NoStop}%
\bibitem [{\citenamefont {Mano}\ \emph {et~al.}(1996)\citenamefont {Mano},
  \citenamefont {Suzuki},\ and\ \citenamefont {Takasugi}}]{Mano:1996vt}%
  \BibitemOpen
  \bibfield  {author} {\bibinfo {author} {\bibfnamefont {S.}~\bibnamefont
  {Mano}}, \bibinfo {author} {\bibfnamefont {H.}~\bibnamefont {Suzuki}}, \ and\
  \bibinfo {author} {\bibfnamefont {E.}~\bibnamefont {Takasugi}},\ }\href
  {\doibase 10.1143/PTP.95.1079} {\bibfield  {journal} {\bibinfo  {journal}
  {Prog. Theor. Phys.}\ }\textbf {\bibinfo {volume} {95}},\ \bibinfo {pages}
  {1079} (\bibinfo {year} {1996})},\ \Eprint
  {http://arxiv.org/abs/gr-qc/9603020} {arXiv:gr-qc/9603020 [gr-qc]}
  \BibitemShut {NoStop}%
\bibitem [{BHP()}]{BHPToolkit}%
  \BibitemOpen
  \href@noop {} {\enquote {\bibinfo {title} {{Black Hole Perturbation
  Toolkit}},}\ }\bibinfo {howpublished}
  {(\href{http://bhptoolkit.org/}{bhptoolkit.org})}\BibitemShut {NoStop}%
\bibitem [{\citenamefont {Poisson}\ and\ \citenamefont
  {Sasaki}(1995)}]{Poisson:1994yf}%
  \BibitemOpen
  \bibfield  {author} {\bibinfo {author} {\bibfnamefont {E.}~\bibnamefont
  {Poisson}}\ and\ \bibinfo {author} {\bibfnamefont {M.}~\bibnamefont
  {Sasaki}},\ }\href {\doibase 10.1103/PhysRevD.51.5753} {\bibfield  {journal}
  {\bibinfo  {journal} {Phys. Rev.}\ }\textbf {\bibinfo {volume} {D51}},\
  \bibinfo {pages} {5753} (\bibinfo {year} {1995})},\ \Eprint
  {http://arxiv.org/abs/gr-qc/9412027} {arXiv:gr-qc/9412027 [gr-qc]}
  \BibitemShut {NoStop}%
\bibitem [{\citenamefont {Yoshino}\ and\ \citenamefont
  {Kodama}(2014)}]{Yoshino:2013ofa}%
  \BibitemOpen
  \bibfield  {author} {\bibinfo {author} {\bibfnamefont {H.}~\bibnamefont
  {Yoshino}}\ and\ \bibinfo {author} {\bibfnamefont {H.}~\bibnamefont
  {Kodama}},\ }\href {\doibase 10.1093/ptep/ptu029} {\bibfield  {journal}
  {\bibinfo  {journal} {PTEP}\ }\textbf {\bibinfo {volume} {2014}},\ \bibinfo
  {pages} {043E02} (\bibinfo {year} {2014})},\ \Eprint
  {http://arxiv.org/abs/1312.2326} {arXiv:1312.2326 [gr-qc]} \BibitemShut
  {NoStop}%
\bibitem [{\citenamefont {Yoshino}\ and\ \citenamefont
  {Kodama}(2015)}]{Yoshino:2015nsa}%
  \BibitemOpen
  \bibfield  {author} {\bibinfo {author} {\bibfnamefont {H.}~\bibnamefont
  {Yoshino}}\ and\ \bibinfo {author} {\bibfnamefont {H.}~\bibnamefont
  {Kodama}},\ }\href {\doibase 10.1088/0264-9381/32/21/214001} {\bibfield
  {journal} {\bibinfo  {journal} {Class. Quant. Grav.}\ }\textbf {\bibinfo
  {volume} {32}},\ \bibinfo {pages} {214001} (\bibinfo {year} {2015})},\
  \Eprint {http://arxiv.org/abs/1505.00714} {arXiv:1505.00714 [gr-qc]}
  \BibitemShut {NoStop}%
\bibitem [{\citenamefont {Ghosh}\ \emph {et~al.}(2019)\citenamefont {Ghosh},
  \citenamefont {Berti}, \citenamefont {Brito},\ and\ \citenamefont
  {Richartz}}]{Ghosh:2018gaw}%
  \BibitemOpen
  \bibfield  {author} {\bibinfo {author} {\bibfnamefont {S.}~\bibnamefont
  {Ghosh}}, \bibinfo {author} {\bibfnamefont {E.}~\bibnamefont {Berti}},
  \bibinfo {author} {\bibfnamefont {R.}~\bibnamefont {Brito}}, \ and\ \bibinfo
  {author} {\bibfnamefont {M.}~\bibnamefont {Richartz}},\ }\href {\doibase
  10.1103/PhysRevD.99.104030} {\bibfield  {journal} {\bibinfo  {journal} {Phys.
  Rev.}\ }\textbf {\bibinfo {volume} {D99}},\ \bibinfo {pages} {104030}
  (\bibinfo {year} {2019})},\ \Eprint {http://arxiv.org/abs/1812.01620}
  {arXiv:1812.01620 [gr-qc]} \BibitemShut {NoStop}%
\bibitem [{\citenamefont {Isi}\ \emph {et~al.}(2019)\citenamefont {Isi},
  \citenamefont {Sun}, \citenamefont {Brito},\ and\ \citenamefont
  {Melatos}}]{Isi:2018pzk}%
  \BibitemOpen
  \bibfield  {author} {\bibinfo {author} {\bibfnamefont {M.}~\bibnamefont
  {Isi}}, \bibinfo {author} {\bibfnamefont {L.}~\bibnamefont {Sun}}, \bibinfo
  {author} {\bibfnamefont {R.}~\bibnamefont {Brito}}, \ and\ \bibinfo {author}
  {\bibfnamefont {A.}~\bibnamefont {Melatos}},\ }\href {\doibase
  10.1103/PhysRevD.99.084042} {\bibfield  {journal} {\bibinfo  {journal} {Phys.
  Rev.}\ }\textbf {\bibinfo {volume} {D99}},\ \bibinfo {pages} {084042}
  (\bibinfo {year} {2019})},\ \Eprint {http://arxiv.org/abs/1810.03812}
  {arXiv:1810.03812 [gr-qc]} \BibitemShut {NoStop}%
\bibitem [{\citenamefont {Goncharov}\ and\ \citenamefont
  {Thrane}(2018)}]{Goncharov:2018ufi}%
  \BibitemOpen
  \bibfield  {author} {\bibinfo {author} {\bibfnamefont {B.}~\bibnamefont
  {Goncharov}}\ and\ \bibinfo {author} {\bibfnamefont {E.}~\bibnamefont
  {Thrane}},\ }\href@noop {} {\  (\bibinfo {year} {2018})},\ \Eprint
  {http://arxiv.org/abs/1805.03761} {arXiv:1805.03761 [astro-ph.IM]}
  \BibitemShut {NoStop}%
\bibitem [{\citenamefont {D'Antonio}\ \emph {et~al.}(2018)\citenamefont
  {D'Antonio} \emph {et~al.}}]{DAntonio:2018sff}%
  \BibitemOpen
  \bibfield  {author} {\bibinfo {author} {\bibfnamefont {S.}~\bibnamefont
  {D'Antonio}} \emph {et~al.},\ }\href {\doibase 10.1103/PhysRevD.98.103017}
  {\bibfield  {journal} {\bibinfo  {journal} {Phys. Rev.}\ }\textbf {\bibinfo
  {volume} {D98}},\ \bibinfo {pages} {103017} (\bibinfo {year} {2018})},\
  \Eprint {http://arxiv.org/abs/1809.07202} {arXiv:1809.07202 [gr-qc]}
  \BibitemShut {NoStop}%
\bibitem [{\citenamefont {Palomba}\ \emph {et~al.}(2019)\citenamefont {Palomba}
  \emph {et~al.}}]{Palomba:2019vxe}%
  \BibitemOpen
  \bibfield  {author} {\bibinfo {author} {\bibfnamefont {C.}~\bibnamefont
  {Palomba}} \emph {et~al.},\ }\href@noop {} {\  (\bibinfo {year} {2019})},\
  \Eprint {http://arxiv.org/abs/1909.08854} {arXiv:1909.08854 [astro-ph.HE]}
  \BibitemShut {NoStop}%
\end{thebibliography}%

\end{document}